\documentclass[11pt]{article}

\usepackage{latexsym, amsthm, amssymb}

\newcommand{\E}{{\textbf{E}}}
\newcommand{\CH}{{\rm ch}}
\newcommand{\poly}{{\rm poly}}

\begin{document}

\title{An Extension of the Lov\'{a}sz Local Lemma, and its Applications to
Integer Programming\thanks{A preliminary version of this work appeared as
a paper of the same title in the 
\textit{Proc.\ ACM-SIAM Symposium on Discrete Algorithms}, 
pages 6--15, 1996. 
Work done in parts at:
(i) the University of Maryland (supported in part by 
NSF Award CCR-0208005), 
(ii) the National University of Singapore,
(iii) DIMACS (supported in part by NSF-STC91-19999 and
by support from the N.J. Commission
on Science and Technology), 
(iv) the Institute for Advanced Study, Princeton, NJ
(supported in part by grant 93-6-6 of the Alfred P. Sloan Foundation),
and 
(v) while visiting the Max-Planck-Institut f\"{u}r Informatik, 
66123 Saarbr\"{u}cken, Germany.}}
\author{Aravind Srinivasan\thanks{Department of Computer Science 
and Institute for
Advanced Computer Studies, University of Maryland, College Park, MD 20742,
USA. E-mail: \texttt{srin@cs.umd.edu}.}}

\date{}
\maketitle

\newcommand{\remove}[1]{}

\newtheorem{theorem}{Theorem}[section]
\newtheorem{lemma}[theorem]{Lemma}
\newtheorem{claim}[theorem]{Claim}
\newtheorem{proposition}[theorem]{Proposition}
\newtheorem{fact}[theorem]{Fact}
\newtheorem{corol}[theorem]{Corollary}
\newtheorem{definition}[theorem]{Definition}
\newtheorem{conjecture}[theorem]{Conjecture}
\newtheorem{remark}[theorem]{Remark}
\newtheorem{assump}[theorem]{Assumption}
\newtheorem{alg}[theorem]{Algorithm}

\newcommand{\yst}{\mbox{$y^{*}$}}
\newcommand{\zo}{\mbox{$\{0,1\}$}}
\newcommand{\zom}{\mbox{$\{0,1\}^m$}}
\newcommand{\zon}{\mbox{$\{0,1\}^n$}}
\newcommand{\zol}{\mbox{$\{0,1\}^{\ell}$}}
\newcommand{\zoi}{\mbox{$\{0,1\}^{i}$}}
\newcommand{\zoj}{\mbox{$\{0,1\}^{j}$}}

\newcommand{\eox}{ \vspace*{1cm} }
\newcommand{\DEF}{\stackrel {\Delta}{=}}
\newcommand{\epr}{\raisebox{-.3ex}{\hspace{.2in}\rule{2mm}{3mm}}}
\newcommand{\polylog}{\mbox{\sl polylog}}
\newcommand{\pram}{\mbox{\sc PRAM}}
\newcommand{\whp}{\mbox\textit{w.h.p.}}
\newcommand{\idest}{\mbox\textit{i.e.,}}
\newcommand{\given}{\mid}

\begin{abstract} 
The Lov\'{a}sz Local Lemma due to Erd\H{o}s and Lov\'{a}sz 
(in \textit{Infinite and Finite Sets}, Colloq.\ Math.\ Soc.\ J.\ Bolyai 11, 
1975, pp.\ 609--627) is a powerful tool in proving the
existence of rare events. We present an extension of this lemma, which
works well when the event to be shown to exist is a
conjunction of individual events, each of which asserts that a random
variable does not deviate much from its mean. As applications,
we consider two
classes of NP-hard integer programs: minimax 
and covering integer programs.
A key technique, \textit{randomized rounding of linear relaxations}, 
was developed by Raghavan \& Thompson 
(\textit{Combinatorica, 7 (1987), pp.\ 365--374})
to derive good approximation
algorithms for such problems. We use
our extension of the Local Lemma to prove that randomized rounding
produces, with non-zero probability, much better feasible solutions than
known before, if the constraint matrices of these integer programs are 
\textit{column-sparse} (e.g., routing using short paths, problems on
hypergraphs with small dimension/degree). This complements certain
well-known results from discrepancy theory. 
We also generalize the \textit{method of pessimistic estimators} due to
Raghavan (\textit{J.\ Computer and System Sciences, 37 (1988), 
pp.\ 130--143}), 
to obtain constructive (algorithmic) versions of our
results for covering integer programs. 
\end{abstract}


\section{Introduction}
\label{section:intro}
The powerful Lov\'asz Local Lemma (LLL) is often 
used to show the existence of \textit{rare} combinatorial structures by
showing that a random sample from a suitable
sample space produces them with positive
probability \cite{erdos-lovasz:lll}; see 
Alon \& Spencer \cite{alon-spencer} and Motwani \& Raghavan
\cite{motwani-raghavan} for 
several such applications. We present an extension of this lemma,
and demonstrate applications to rounding fractional solutions for 
certain families of integer programs. 

Let $e$ denote the base of natural logarithms as usual.
The symmetric case of the LLL shows that all
of a set of ``bad'' events $E_i$ can be avoided under some conditions:

\begin{lemma}
\label{lemma:lll}
{\rm \textbf{(\cite{erdos-lovasz:lll})}}
Let $E_1, E_2, \ldots, E_m$ be any events with $\Pr(E_i) \leq p ~ \forall i$.
If each $E_i$ is mutually independent of all but
at most $d$ of the other events $E_j$ and if $ep(d+1) \leq 1$, then
$\Pr(\bigwedge_{i=1}^{m} \overline{E_i}) > 0$.
\end{lemma}

Though the LLL is powerful, one problem is that the ``dependency'' $d$
is high in some cases, precluding the use of the LLL if $p$ is
not small enough. We present a partial solution to this via
an extension of the LLL (Theorem~\ref{theorem:lll-new}), which shows how to 
essentially reduce $d$ for a class of
events $E_i$; this works well when each $E_i$ denotes some random 
variable deviating ``much'' from its mean. In a nutshell, we show that
such events $E_i$ can often be decomposed suitably into sub-events; 
although the sub-events may have a large dependency among themselves,
we show that it suffices to have a small ``bipartite dependency'' between
the set of events $E_i$ and the set of sub-events. This, in combination
with some other ideas, leads to the following applications in
integer programming. 

It is well-known that a large number of NP-hard combinatorial
optimization problems can be cast as integer linear programming
problems (ILPs). Due to their NP-hardness, good approximation algorithms
are of much interest for such problems. Recall that a 
$\rho$-approximation algorithm for a minimization problem 
is a polynomial-time algorithm that delivers a solution whose
objective function value is at most $\rho$ times optimal; $\rho$
is usually called the \textit{approximation guarantee}, 
\textit{approximation ratio}, or \textit{performance guarantee} of
the algorithm. Algorithmic work in this area typically focuses on
achieving the smallest possible $\rho$ in polynomial time. One
powerful paradigm here is to start with the 
linear programming (LP) relaxation of the given ILP wherein the variables
are allowed to be \textit{reals} within their integer ranges; once
an optimal solution is found for the LP, the main issue is how to
\textit{round} it to a good feasible solution for the ILP. 
Rounding results in this context often have the following 
strong property: they
present an integral solution of value at most $\yst \cdot \rho$,
where $\yst$ will throughout denote the optimal solution
value of the LP relaxation. Since the optimal solution value $OPT$ of the ILP
is easily seen to be lower-bounded by $\yst$, such rounding algorithms
are also $\rho$-approximation algorithms. Furthermore, they provide an
upper bound of $\rho$ on the ratio $OPT/\yst$, which is usually called
the \textit{integrality gap} or \textit{integrality ratio} of the
relaxation; the smaller this value, the better the relaxation. 

This work presents improved upper bounds on the integrality gap of
the natural LP relaxation for two families of ILPs: minimax 
integer programs (MIPs) and covering integer programs (CIPs). 
(The precise definitions and results are presented in 
\S~\ref{section:ip-defs}.) For the latter,
we also provide the corresponding polynomial-time rounding algorithms.
Our main improvements are in the case where the coefficient
matrix of the given ILP is \textit{column-sparse}: i.e., the number of
nonzero entries in every column is bounded by a given parameter $a$.
There are classical rounding theorems for such column-sparse problems
(e.g., Beck \& Fiala \cite{beck-fiala}, Karp, Leighton, Rivest, Thompson, 
Vazirani \& Vazirani \cite{klrtvv}). Our results complement, and are
incomparable with, these results. Furthermore, the notion of
column-sparsity, which denotes no variable occurring in ``too many''
constraints, occurs naturally in combinatorial optimization: 
e.g., routing using ``short'' paths, and
problems on hypergraphs with ``small'' degree. These issues are
discussed further in \S~\ref{section:ip-defs}. 

A key technique, \textit{randomized rounding of linear relaxations}, was
developed by Raghavan \& Thompson \cite{raghavan-thompson} to get 
approximation algorithms for such ILPs.
We use Theorem~\ref{theorem:lll-new} to prove that this technique
produces, with non-zero probability, much better feasible solutions than
known before, if the constraint matrix of the given MIP/CIP is
column-sparse. (In the case of MIPs, our algorithm iterates 
randomized rounding several times with different choices of 
parameters, in order to achieve our result.) 
Such results cannot be got via Lemma~\ref{lemma:lll},
as the dependency $d$, in the sense of Lemma~\ref{lemma:lll}, can 
be as high as $\Theta(m)$ for these problems. Roughly speaking, 
Theorem~\ref{theorem:lll-new} helps show that if no column in our 
given ILP has more than $a$ nonzero entries, then the dependency 
can essentially be brought down to a polynomial in $a$; this is the
key driver behind our improvements. 

Theorem~\ref{theorem:lll-new} works well in
combination with an idea that has blossomed in the areas of 
derandomization and pseudorandomness, in the last two decades: (approximately)
decomposing a function of several variables into a sum of terms, each of
which depends on only a few of these variables. Concretely, suppose $Z$ is 
a sum of random variables $Z_i$. Many tools have been developed
to upper-bound $\Pr(Z - \E[Z] \geq z)$ and
$\Pr(|Z - \E[Z]| \geq z)$ even if the $Z_i$s are
only (almost) $k$-wise independent for some ``small'' $k$, 
rather than completely
independent. The idea is to bound the probabilities by considering
$\E[(Z - \E[Z])^k]$ or similar expectations, which look at the $Z_i$ 
\textit{$k$ or fewer at a time} (via linearity of 
expectation). The main application of this has been that the $Z_i$ can then
be sampled using ``few'' random bits, yielding a 
derandomization/pseudorandomness result (e.g.,
\cite{abi,luby:mis,berger-rompel,motwani-naor-naor,nisan:pseudo,sss}). 
Our results show that such ideas can in fact be used to show that some 
structures exist! This is one of our main contributions.

What about polynomial-time algorithms for our existential results?
Typical applications of Lemma~\ref{lemma:lll} are ``nonconstructive''
[i.e., do not directly imply (randomized) polynomial-time
algorithmic versions], since the positive probability guaranteed
by Lemma~\ref{lemma:lll} can be exponentially small in the size of
the input. However, certain algorithmic versions of the LLL
have been developed
starting with the seminal work of Beck \cite{beck:lll}. These ideas
do not seem to apply to our extension of the LLL, and hence our
MIP result is nonconstructive. Following the preliminary version
of this work \cite{srin:lll}, two main algorithmic versions related
to our work have been obtained: (i) for a subclass of the MIPs 
\cite{llrs}, and (ii) for a somewhat different notion of approximation
than the one we study, for certain families of MIPs \cite{czu-schei:lll2}.

Our main algorithmic contribution is for CIPs and multi-criteria
versions thereof: we show, by a 
generalization of the \textit{method of pessimistic estimators} 
\cite{raghavan:rounding}, that we can efficiently construct
the same structure as is guaranteed by our nonconstructive
argument. We view this as interesting for two reasons. First, the
generalized pessimistic estimator argument requires a quite delicate
analysis, which we expect to be useful in other applications of developing
constructive versions of existential arguments. Second, 
except for some of the algorithmic versions of the LLL developed in
\cite{mr:lll,molloy-reed:book}, most current
algorithmic versions minimally require something like
``$p d^3 = O(1)$'' (see, e.g., \cite{beck:lll,alon:lll}); 
the LLL only needs that $p d = O(1)$. While
this issue does not matter much in many applications, it crucially does,
in some others. A good example of this is the
existentially-optimal integrality gap for the edge-disjoint
paths problem with ``short'' paths, shown using the LLL in 
\cite{lrs:mcf}. The above-seen ``$p d^3 = O(1)$'' requirement of
currently-known algorithmic approaches to the LLL, leads to algorithms
that will violate the edge-disjointness condition when applied in
this context: specifically, they may route up to three paths on some
edges of the graph. 
See \cite{bfu:static-dynamic} for a different -- random-walk based --
approach to low-congestion routing. An algorithmic version
of this edge-disjoint paths result of \cite{lrs:mcf} is
still lacking. It is a very interesting open question
whether there is an algorithmic version of the LLL that can construct
the same structures as guaranteed to exist by the LLL. In particular,
can one of the most successful derandomization tools -- the method
of conditional probabilities or its generalization, 
the pessimistic estimators method -- be applied, fixing the underlying
random choices of the probabilistic argument one-by-one? This intriguing
question is open (and seems difficult) for now. As a step in this
direction, we are able to show how such approaches can indeed be developed,
in the context of CIPs. 

Thus, our main contributions are as follows.
(a) The LLL extension is of independent interest: it helps in 
certain settings where the ``dependency'' among the ``bad'' events 
is too high for the LLL to be directly applicable. We expect to see
further applications/extensions of such ideas. 
(b) This work shows that certain classes of column-sparse ILPs have 
much better solutions than known before; such problems abound in practice
(e.g., short paths are often desired/required in routing). 
(c) Our generalized method of pessimistic estimators should
prove fruitful in other contexts also; it is a step toward complete
algorithmic versions of the LLL. 

The rest of this paper is organized as follows. Our results are
first presented in \S~\ref{section:ip-defs}, along with a 
discussion of related work. The extended LLL, and some large-deviation
methods that will be seen to work well with it, 
are shown in \S~\ref{section:lll-ext}. 
Sections~\ref{section:mip} and~\ref{section:cip} are devoted to our
rounding applications. Finally, \S~\ref{section:concl} concludes. 

\section{Our Results and Related Work}
\label{section:ip-defs}
Let $Z_+$ denote the set of non-negative integers; 
for any $k \in Z_+$, $[k] \doteq 
\{1, \ldots, k\}$. ``Random variable'' is abbreviated by ``r.v.'', and
logarithms are to the base $2$ unless specified otherwise. 

\begin{definition}
\label{defn:mip}
\textbf{(Minimax Integer Programs)}
An MIP (minimax integer program) has variables $W$ and
$\{x_{i,j}: i \in [n], j \in [\ell_i] \}$, for some integers $\{\ell_i\}$.
Let $N = \sum_{i \in [n]} \ell_i$ and let $x$ denote the $N$-dimensional vector
of the variables $x_{i,j}$ (arranged in any fixed order). An MIP seeks to
minimize $W$, an unconstrained real, subject to: 
\begin{itemize}
\item[(i)] 
Equality constraints: $\forall i \in [n] ~ \sum_{j \in [\ell_i]} x_{i,j}
= 1$; 
\item[(ii)] a system of linear inequalities $Ax \leq \vec{W}$, where
$A \in [0,1]^{m \times N}$ and $\vec{W}$ is the $m$-dimensional vector
with the variable $W$ in each component, and
\item[(iii)] Integrality constraints: $x_{i,j} \in \zo ~ \forall i,j$.
\end{itemize}
We let $g$ denote the maximum column sum in any column of $A$, and 
$a$ be the maximum number of non-zero entries in any column of $A$.
\end{definition}

To see what problems MIPs model, note, from constraints (i) and 
(iii) of MIPs, that for all $i$, any feasible solution will
make the set $\{x_{i,j}: j \in [\ell_i]\}$ have precisely one 1, with all
other elements being 0; MIPs thus model many ``choice'' scenarios. 
Consider, e.g., global routing in VLSI 
gate arrays \cite{raghavan-thompson}.
Given are an undirected graph $G = (V,E)$, 
a function $\rho: V \rightarrow V$, and
$\forall i \in V$, a set $P_i$ of paths in $G$, each connecting
$i$ to $\rho(i)$; we must connect each $i$ with $\rho(i)$ 
using exactly
one path from $P_i$, so that the maximum number of times that any
edge in $G$ is used for, is minimized--an MIP formulation is obvious,
with $x_{i,j}$ being the indicator variable for
picking the $j$th path in $P_i$. 
This problem, the vector-selection 
problem of \cite{raghavan-thompson}, and the discrepancy-type
problems of Section~\ref{section:mip}, are
all modeled by MIPs; many MIP instances, e.g., global routing,
are NP-hard.

We now introduce the next family of integer programs that we will work
with. 

\begin{definition}
\label{defn:cip}
\textbf{(Covering Integer Programs)}
Given $A \in [0,1]^{m \times n}$, $b \in [1, \infty)^m$ 
and $c \in [0,1]^n$ with max$_j~ c_j = 1$, 
a covering integer program (CIP) seeks to minimize $c^T \cdot x$ subject
to $x \in Z_{+}^n$ and $Ax \geq b$.
If  $A \in \{0,1\}^{m \times n}$, 
each entry of $b$ is assumed integral. 
We define $B = \min_i ~ b_i$, and let $a$ be the maximum 
number of non-zero entries in any column of $A$. A CIP is called
\textit{unweighted} if $c_j = 1$ $\forall j$, and \textit{weighted} 
otherwise.
\end{definition}

Note the parameters $g$, $a$ and $B$ of definitions \ref{defn:mip} 
and \ref{defn:cip}. Though there are usually no restrictions on the 
entries of $A, b$ and $c$ in CIPs aside of
non-negativity, it is well-known and easy to check that 
the above restrictions are without loss of generality.
CIPs again model many NP-hard problems in 
combinatorial optimization. Recall that a hypergraph
$H = (V,E)$ is a family of subsets $E$ (edges)
of a set $V$ (vertices). The classical
\textit{set cover} problem--covering
$V$ using the smallest number of edges in $E$ (and its natural
weighted version) is a standard example of a CIP. 
The parameter $a$ here is the maximum number of vertices in
any edge. 

Next, there is growing interest in \textit{multi-criteria}
optimization, since different participating individuals
and/or organizations
may have different objective functions in a given problem instance;
see, e.g., \cite{pap-yan:multi-crit}. Motivated by this, we 
study multi-criteria optimization in the setting of covering problems: 

\begin{definition}
\label{defn:multi-cip}
\textbf{(Multi-criteria CIPs; informal)}
A \textit{multi-criteria} CIP has a system of constraints
``$Ax \geq b$'' as in CIPs, and has \textit{several} given
non-negative vectors $c_1, c_2, \ldots, c_{\ell}$; the aim 
is to keep \textit{all} the values $c_i^T \cdot x$ ``low''. 
(For instance, we may aim to minimize $\max_i c_i^T \cdot x$
subject to $Ax \geq b$.) As in Definition~\ref{defn:cip},
we assume that $A \in [0,1]^{m \times n}$, $b \in [1, \infty)^m$ 
and for all $i$, $c_i \in [0,1]^n$ with max$_j~ c_{i,j} = 1$.
\end{definition}

We now present a lemma to quantify our
approximation results; its proof is given in \S \ref{section:lll-ext}. 

\begin{lemma}
\label{lemma:G-bounds}
Given independent r.v.s $X_1, \ldots , X_n \in [0,1]$,
let $X = \sum_{i=1}^{n} X_i$ and $\mu = \E[X]$. 
\begin{itemize}
\item[a.] For any $\delta > 0$, 
$\Pr(X \geq \mu(1 + \delta)) \leq G(\mu, \delta)$, where
$G(\mu, \delta) = 
\left(e^{\delta}/(1 + \delta)^{1 + \delta}\right)^{\mu}$. 
\item[b.] $\forall \mu > 0 ~ \forall p \in (0,1),~
\exists \delta = H(\mu, p) > 0$ such that
$\lceil \mu \delta \rceil \cdot G(\mu, \delta) \leq p$ and such that
\[ H(\mu,p) = 
\Theta\left(\frac{\log(p^{-1})}{\mu \log(\log(p^{-1}) / \mu)}\right) 
\mbox{ if $\mu \leq  \log(p^{-1})/2$;~~} 
H(\mu,p) = \Theta\left(\sqrt{\frac{\log(\mu + p^{-1})}{\mu}}\right) 
\mbox{ otherwise.} \]
\end{itemize}
\end{lemma}

Given an ILP, we can find an optimal solution $x^*$ to its
LP relaxation efficiently, but
need to round fractional entries in $x^*$ to integers. The idea of
randomized rounding is: given a real $v > 0$, 
round $v$ to $\lfloor v \rfloor + 1$ with probability 
$v - \lfloor v \rfloor$, and round $v$ to $\lfloor v \rfloor$
with probability $1 - v + \lfloor v \rfloor$. This has the nice property
that the mean outcome is $v$. Starting with this idea, the analysis of 
\cite{raghavan-thompson} produces an integral solution of value at most 
$\yst + O(\min\{\yst,m\} \cdot H(\min\{\yst,m\},1/m))$
for MIPs (though phrased a bit differently); this is
derandomized in \cite{raghavan:rounding}. But this does not exploit
the sparsity of $A$; the previously-mentioned result of 
\cite{klrtvv} produces an
integral solution of value at most $\yst + g + 1$. 

For CIPs, the idea is to solve the 
LP relaxation, scale \textit{up} the components of $x^*$ suitably, and
then perform randomized rounding; see 
Section~\ref{section:cip} for the details. Starting with this
idea, the work of \cite{raghavan-thompson} leads to certain approximation
bounds; similar bounds are achieved through different means by
Plotkin, Shmoys \& Tardos \cite{pst}. Work of this author 
\cite{srin:pos-correl} improved upon these
results by observing a ``correlation'' property of CIPs,
getting an approximation ratio of 
$1 + O(\mbox{max}\{\ln(mB/\yst)/B, \sqrt{\ln(mB/\yst)/B}\})$. 
Thus, while the work of \cite{raghavan-thompson} gives a general 
approximation bound for MIPs, the result of \cite{klrtvv}
gives good results for sparse MIPs. For CIPs, the current-best
results are those of \cite{srin:pos-correl}; however, no better results
were known for sparse CIPs.

\subsection{Improvements achieved}
\label{section:improvements}
For MIPs, we use the extended LLL and an idea of \'{E}va Tardos that leads
to a bootstrapping of the LLL extension, to show the existence of an 
integral solution of value
$\yst + O(\min\{\yst,m\} \cdot 
H(\min\{\yst,m\},1/a)) + O(1)$; see Theorem~\ref{theorem:mip-bootstrap}. 
Since $a \leq m$, this is always as good as the
$\yst + O(\min\{\yst,m\} \cdot H(\min\{\yst,m\},1/m))$
bound of \cite{raghavan-thompson} and is a good improvement, if
$a \ll m$. It also is
an improvement over the additive $g$ factor of \cite{klrtvv} in
cases where $g$ is not small compared to $\yst$. 

Consider, e.g., the global routing problem and its MIP
formulation, sketched above; $m$ here is the number of edges in
$G$, and $g=a$ is the maximum length of any path in $\bigcup_i P_i$.
To focus on a specific interesting case, suppose $\yst$,
the fractional congestion, is at most one. Then while the 
previous results (\cite{raghavan-thompson} and \cite{klrtvv}, resp.)
give bounds of $O(\log m / \log\log m)$ and $O(a)$ on
an integral solution, we get the improved bound of $O(\log a /\log\log a)$.
Similar improvements are easily seen for
other ranges of $\yst$ also; e.g., if $\yst = O(\log a)$, an integral
solution of value $O(\log a)$ exists, improving on the previously
known bounds of $O(\log m / \log(2 \log m / \log a))$ and $O(a)$.
Thus, routing along \textit{short} paths (this is the
notion of sparsity for the global routing problem) is very beneficial
in keeping the congestion low. Section~\ref{section:mip} 
presents a scenario where we get such improvements, for discrepancy-type 
problems \cite{spencer,alon-spencer}. In particular, we 
generalize a hypergraph-partitioning 
result of F\"{u}redi \& Kahn \cite{furedi-kahn}.

Recall the bounds of \cite{srin:pos-correl} for CIPs mentioned
in the paragraph preceding this subsection; 
our bounds for CIPs depend only on the set of constraints
$Ax \geq b$, i.e., they hold for any non-negative
objective-function vector $c$. Our improvements over \cite{srin:pos-correl}
get better as $\yst$ decreases. 
We show an integrality gap of
$1 + O(\mbox{max}\{\ln (a+1) /B, \sqrt{\ln (a+1)/B}\})$, 
once again improving on \cite{srin:pos-correl} for \textit{weighted} CIPs. 
This CIP
bound is better than that of \cite{srin:pos-correl} if $\yst \leq mB/a$: 
this inequality fails for unweighted CIPs and is generally true
for weighted CIPs, since $\yst$ can get arbitrarily small in the latter case.
In particular, we generalize the result of 
Chv\'{a}tal \cite{chvatal:covers} on weighted set cover. 
Consider, e.g., a
facility location problem on a directed graph $G = (V,A)$: given a cost
$c_i \in [0,1]$ for each $i \in V$, we want a min-cost assignment of
facilities to the nodes such that each node sees at least $B$ facilities
in its out-neighborhood--multiple facilities at a node are allowed. If
$\Delta_{in}$ is the maximum in-degree of $G$, we show an integrality
gap of 
$1 + O(\mbox{max}\{\ln(\Delta_{in} +1)/B, 
\sqrt{\ln(B(\Delta_{in}+1))/B}\})$. 
This improves on \cite{srin:pos-correl} if $\yst \leq |V|B/\Delta_{in}$;
it shows an $O(1)$ 
(resp., $1 + o(1)$) integrality gap if $B$ grows as fast as (resp., strictly
faster than) $\log \Delta_{in}$. Theorem~\ref{theorem:cip} presents our
covering results. 

A key corollary of our results is that for families of instances of
CIPs, we get a good ($O(1)$ or $1 + o(1)$) 
integrality gap 
if $B$ grows at least as fast as $\log a$. 
Bounds on the result of a 
greedy algorithm for CIPs relative to the optimal \textit{integral} solution,
are known \cite{dobson,fisher-wolsey}.
Our bound improves that of \cite{dobson} and is incomparable with 
\cite{fisher-wolsey}; for any given $A$, $c$, and the
unit vector $b / ||b||_2$, our bound improves on \cite{fisher-wolsey} if
$B$ is more than a certain threshold. 
As it stands, randomized rounding produces such improved solutions
for several CIPs only with a very low,
sometimes exponentially small, probability. Thus, it does not
imply a randomized algorithm, often. To this end, we generalize 
Raghavan's method of pessimistic estimators to derive an algorithmic
(polynomial-time) version of our results for CIPs, in 
\S~\ref{section:construct}.

We also show via Theorem~\ref{theorem:mult-cip} and 
Corollary~\ref{cor:mult-cip} that multi-criteria CIPs can be
approximated well. In particular, Corollary~\ref{cor:mult-cip}
shows some interesting cases where the approximation guarantee
for multi-criteria CIPs grows in a very much sub-linear fashion
with the number $\ell$ of given vectors $c_i$: 
the approximation ratio is at most $O(\log \log \ell)$ times
what we show for CIPs (which correspond to the case where
$\ell=1$). We are not aware of any such earlier work on
multi-criteria CIPs. 

The preliminary version of this work was presented in
\cite{srin:lll}. As mentioned in \S~\ref{section:intro},
two main algorithmic versions related to our work have been obtained
following \cite{srin:lll}. First, for a subclass of the MIPs where the
nonzero entries of the matrix $A$ are ``reasonably large'', constructive
versions of our results have been obtained in \cite{llrs}. 
Second, for a notion of approximation that is different from
the one we study, algorithmic results have been developed for
certain families of MIPs in \cite{czu-schei:lll2}.
Furthermore, our Theorem~\ref{theorem:cip} for CIPs has been
used in \cite{koll-young} to develop approximation
algorithms for CIPs that have given upper bounds on the variables $x_j$. 

\section{The Extended LLL and an Approach to Large Deviations}
\label{section:lll-ext}
We now present our LLL extension, Theorem~\ref{theorem:lll-new}.
For any event $E$, define $\chi(E)$ to be its indicator r.v.: 
$1$ if $E$ holds and $0$ otherwise. 
Suppose we have ``bad'' events $E_1, \ldots, E_m$ with a ``dependency''
$d'$ (in the sense of Lemma~\ref{lemma:lll}) that is ``large''. 
Theorem~\ref{theorem:lll-new} shows how to essentially replace $d'$ by
a possibly much-smaller $d$, under some conditions. 
It generalizes Lemma~\ref{lemma:lll} (define one r.v., 
$C_{i,1} = \chi(E_i)$,
for each $i$, to get Lemma~\ref{lemma:lll}), its proof
is very similar to the classical proof of Lemma \ref{lemma:lll}, 
and its motivation will be clarified by the applications.

\begin{theorem}
\label{theorem:lll-new}
Given events $E_1,\ldots,E_m$ and any $I \subseteq [m]$,
let $Z(I) \doteq \bigwedge_{i\in I}\overline{E_i}$. 
Suppose that for some positive integer $d$, we can define,
for each $i \in [m]$, a finite number of 
r.v.s $C_{i,1},C_{i,2},\ldots$ each taking on only \textit{non-negative} values
such that:
\begin{itemize}
\item[(i)] any $C_{i,j}$ is mutually independent of all but
at most $d$ of the events $E_k$, $k \not= i$, and 
\item[(ii)] $\forall I \subseteq ([m] - \{i\})$,
$\Pr(E_i \bigm| Z(I)) \leq \sum_j \E[C_{i,j} \bigm| Z(I)]$. 
\vspace*{-0.06in}
\end{itemize}
Let $p_i$ denote
$\sum_j \E[C_{i,j}]$; clearly,
$\Pr(E_i) \leq p_i$ (set $I = \phi$ in (ii)). Suppose that
for all $i \in [m]$ we have $ep_i (d + 1) \leq 1$.
Then $\Pr(\bigwedge_i \overline{E_i}) \geq (d/(d+1))^m > 0$.
\end{theorem}
\vspace*{0.1in}

\noindent 
\begin{remark}
$C_{i,j}$ and $C_{i,j'}$ can ``depend''
on \textit{different} subsets of $\{E_k | k \not= i \}$; the only 
restriction is that these subsets be of size at most $d$. 
Note that we have essentially reduced the dependency among the
$E_i$s, to just $d$: $e p_i(d+1) \leq 1$ suffices. Another 
important point 
is that the dependency \textit{among the r.v.s $C_{i,j}$} could be
much higher than $d$: all we count is the number of $E_k$ that any
$C_{i,j}$ depends on. 
\end{remark}

\vspace*{0.1in}

\noindent 
\textit{Proof of Theorem~\ref{theorem:lll-new}.} 
We prove by induction on $|I|$
that if $i\not\in I$ then $\Pr(E_i \bigm| Z(I)) \leq ep_i$, which suffices
to prove the theorem since $\Pr(\bigwedge_i \overline{E_i}) =
\prod_{i \in [m]} (1 - \Pr(E_i \bigm| Z([i-1])))$.
For the base case where $I=\emptyset$, $\Pr(E_i \bigm| Z(I)) = \Pr(E_i) 
\leq p_i$. For the inductive step, let
$S_{i,j,I} \doteq \{k \in I \bigm| \hbox{$C_{i,j}$ depends
on $E_k$}\}$, and $S'_{i,j,I} = I - S_{i,j,I}$; note that 
$|S_{i,j,I}| \leq d$. If $S_{i,j,I} = \emptyset$, then
$\E[C_{i,j} \bigm| Z(I)] = \E[C_{i,j}]$. Otherwise, 
letting $S_{i,j,I} = \{\ell_1, \ldots, \ell_r\}$, we have
\[ \E[C_{i,j} \bigm| Z(I)] = 
\frac{\E[C_{i,j} \cdot \chi(Z(S_{i,j,I})) \bigm| Z(S'_{i,j,I})]}
{\Pr(Z(S_{i,j,I}) \bigm| Z(S'_{i,j,I}))} 
\leq \frac{\E[C_{i,j} \bigm| Z(S'_{i,j,I})]} 
{\Pr(Z(S_{i,j,I}) \bigm| Z(S'_{i,j,I}))}, \]
since $C_{i,j}$ is non-negative.
The numerator of the last term is $\E[C_{i,j}]$, by assumption. 
The denominator can be lower-bounded as follows:
\[\prod_{s\in [r]}(1 - \Pr(E_{\ell_s} \bigm| Z(\{\ell_1, \ell_2, \ldots, 
\ell_{s-1}\} \cup S'_{i,j,I}))) \geq \prod_{s \in [r]}(1 - ep_{\ell_s}) 
\geq (1 - 1/(d+1))^r \geq (d/(d+1))^d > 1/e; \]
the first inequality follows from the induction hypothesis.
Hence, $\E[C_{i,j} \bigm| Z(I)] \leq e \E[C_{i,j}]$ and thus,
$\Pr(E_i \bigm| Z(I)) \leq \sum_j \E[C_{i,j} \bigm| Z(I)] \leq e p_i
\leq 1/(d+1)$. \qed

\smallskip
The crucial point is that the events $E_i$ could have a
large dependency $d'$, in the sense of the classical 
Lemma~\ref{lemma:lll}. The main utility of Theorem~\ref{theorem:lll-new}
is that if we can ``decompose'' each $E_i$ into the
r.v.s $C_{i,j}$ that satisfy the conditions of the theorem, then
there is the possibility of effectively reducing the dependency by 
much ($d'$ can be replaced by the value $d$). Concrete instances of
this will be studied in later sections.

The tools behind our MIP application are our new LLL, and a result 
of \cite{sss}. Define, for $z = (z_1, \ldots , z_n) \in
\Re^n$, a family of polynomials $S_j(z), j = 0,
1, \ldots , n$, where $S_0(z) \equiv 1$, and for $j \in [n]$,
\begin{equation}
\label{eqn:S_j}
S_j(z) \doteq \sum_{1 \leq i_1 < i_2 \cdots < i_j \leq n}
z_{i_1}z_{i_2} \cdots z_{i_j}.
\end{equation}

\begin{remark}
\label{remark:facto}
For real $x$ and non-negative integral $r$, we define
${x \choose r} \doteq x(x-1) \cdots (x-r+1)/r!$ as usual; this is
the sense meant in Theorem~\ref{theorem:sss} below.
\end{remark}

We define a nonempty event to be any event with a nonzero probability
of occurrence. The relevant theorem of \cite{sss} is the following: 

\begin{theorem} 
\label{theorem:sss}
{\rm \textbf{(\cite{sss})}}
Given r.v.s $X_1, \ldots , X_n \in [0,1]$,
let $X = \sum_{i=1}^{n} X_i$ and $\mu = \E[X]$. Then,

\begin{itemize}
\item[(a)] For any $q > 0$, any
nonempty event $Z$ and any non-negative integer $k \leq q$,
\[ \Pr(X \geq q \bigm| Z) \leq \E[Y_{k,q} \bigm| Z], \]
where $ Y_{k,q} = S_k(X_1, \ldots , X_n)/{q \choose k}$.

\item[(b)] If the $X_i$s are independent, $\delta > 0$, and
$k = \lceil \mu \delta \rceil$,
then $\Pr(X \geq \mu(1 + \delta)) \leq \E[Y_{k,\mu(1 + \delta)}]
\leq G(\mu, \delta)$, where
$G(\cdot, \cdot)$ is as in Lemma~\ref{lemma:G-bounds}. 

\item[(c)] If the $X_i$s are independent, then
$\E[S_k(X_1, \ldots , X_n)] \leq {n \choose k} \cdot (\mu/n)^k
\leq \mu^k/k!$. 
\end{itemize}
\end{theorem}

\begin{proof}
Suppose $r_1, r_2, \ldots r_n \in [0,1]$ satisfy 
$\sum_{i=1}^{n} r_i \geq q$. Then, a simple proof is given
in \cite{sss}, for the fact that for any non-negative integer
$k \leq q$, $S_k(r_1, r_2, \ldots , r_n) \geq {q \choose k}$.
This clearly holds even
given the occurrence of any nonempty event $Z$. Thus we get
$\Pr(X \geq q) \bigm| Z) \leq \Pr(Y_{k,q} \geq 1 \bigm| Z) 
\leq \E[Y_{k,q} \bigm| Z]$,
where the second inequality follows from Markov's inequality. The proofs
of (b) and (c) are given in \cite{sss}.
\end{proof}

We next present the proof of Lemma~\ref{lemma:G-bounds}:

\vspace*{0.1in}

\noindent 
\textit{Proof of Lemma~\ref{lemma:G-bounds}.}
Part (a) is the Chernoff-Hoeffding bound (see, e.g., Appendix A of 
\cite{alon-spencer}, or \cite{motwani-raghavan}). 
For (b), we proceed as follows. 
For any $\mu > 0$, it is easy to check that
\begin{eqnarray}
G(\mu, \delta) & = & 
e^{-\Theta(\mu\delta^2)} ~~\mbox{if $\delta \in (0,1)$; }
\label{eqn:G-behav1} \\ 
G(\mu, \delta) & = & e^{-\Theta(\mu(1+\delta)\ln(1 + \delta))} ~~\mbox{if
$\delta \geq 1$}.
\label{eqn:G-behav2}
\end{eqnarray}
Now if $\mu \leq  \log(p^{-1})/2$, choose 
\[ \delta = C \cdot \frac{\log(p^{-1})}{\mu \log(\log(p^{-1}) / \mu)} \]
for a suitably large constant $C$. Note that $\delta$ is lower-bounded
by some positive constant; hence, (\ref{eqn:G-behav2}) holds 
(since the constant $1$ in the conditions 
``$\delta \in (0,1)$'' and ``$\delta > 1$''
of (\ref{eqn:G-behav1}) and (\ref{eqn:G-behav2}) can clearly
be replaced by any other positive constant). 
Simple algebraic manipulation now shows that if $C$ is large enough,
then $\lceil \mu \delta \rceil \cdot G(\mu, \delta) \leq p$ holds.
Similarly, if $\mu > \log(p^{-1})/2$, we set 
$\delta = C \cdot \sqrt{\frac{\log(\mu + p^{-1})}{\mu}}$ for a large
enough constant $C$, and use (\ref{eqn:G-behav1}).
\qed

\section{Approximating Minimax Integer Programs}
\label{section:mip}
Suppose we are given an MIP conforming to Definition~\ref{defn:mip}.
Define $t$ to be 
$\mbox{max}_{i \in [m]} NZ_i$, where $NZ_i$ is the number of rows of $A$
which have a non-zero coefficient corresponding to at least one variable among
$\{x_{i,j}: j \in [\ell_i]\}$. Note that
\begin{equation}
\label{eqn:mip-ineq}
g \leq a \leq t \leq \min\{m, a \cdot 
\mbox{max}_{i \in [n]} \ell_i \}.
\end{equation}
Theorem~\ref{theorem:mip} now shows how Theorem~\ref{theorem:lll-new} 
can help, for sparse MIPs--those where $t \ll m$. We will then bootstrap
Theorem~\ref{theorem:mip} to get the further improved
Theorem~\ref{theorem:mip-bootstrap}. We start with a 
proposition, whose proof is a simple calculus exercise: 

\begin{proposition}
\label{prop:G-simple}
If $0 < \mu_1 \leq \mu_2$,
then for any $\delta > 0$, $G(\mu_1, \mu_2 \delta/\mu_1) \leq
G(\mu_2, \delta)$. 
\end{proposition}

\begin{theorem}
\label{theorem:mip}
Given an MIP conforming to Definition~\ref{defn:mip}, randomized rounding
produces a feasible solution of value at most $\yst + 
\lceil \min\{\yst, m\} \cdot H(\min\{\yst,m\}, 1/(et)) \rceil$,
with non-zero probability.
\end{theorem}
\begin{proof}
Conduct randomized rounding: 
independently for each $i$, randomly round exactly one $x_{i,j}$ to 
1, guided by the ``probabilities'' $\{x_{i,j}^*\}$. 
We may assume that $\{x_{i,j}^*\}$ is a \textit{basic} feasible solution
to the LP relaxation. Hence, at most $m$ of the $\{x_{i,j}^*\}$ will be
neither zero nor one, and only these variables will participate in the
rounding. Thus, since all the entries of $A$ are in $[0,1]$, we assume
without loss of generality 
from now on that $\yst \leq m$ (and that $\mbox{max}_{i \in [n]}
\ell_i \leq m$); this explains the ``$\min\{\yst, m\}$'' term in
our stated bounds. 
If $z \in \{0,1\}^N$ denotes the
randomly rounded vector, then $\E[(Az)_i] = b_i$ 
by linearity of expectation, i.e., at most $\yst$. Defining 
$k = \lceil \yst H(\yst, 1/(et)) \rceil$ and events
$E_1, E_2, \ldots, E_m$ by $E_i \equiv 
``(Az)_i \geq b_i + k$'', we now show that
$\Pr(\bigwedge_{i \in [m]} \overline{E_i}) > 0$ using 
Theorem~\ref{theorem:lll-new}. Rewrite the $i$th constraint of the MIP as 
\[ \sum_{r \in [n]} X_{i,r} \leq W,
\mbox{ where } 
X_{i,r} = \sum_{s \in [\ell_r]} A_{i,(r,s)} x_{r,s}; \]
the notation $A_{i,(r,s)}$ assumes that the pairs $\{(r,s): r \in [n], 
s \in [\ell_r]\}$ have been mapped bijectively to $[N]$, in some fixed way.
Defining the r.v. 
\[ Z_{i,r} = \sum_{s \in [\ell_r]} A_{i,(r,s)} z_{r,s}, \]
we note that for each $i$, the r.v.s $\{Z_{i,r}: r \in [n]\}$ lie in
$[0,1]$ and are \textit{independent}. Also, 
$E_i \equiv ``\sum_{r \in [n]} Z_{i,r} \geq b_i + k$''.

Theorem~\ref{theorem:sss} suggests a suitable choice for the crucial
r.v.s $C_{i,j}$ (to apply Theorem~\ref{theorem:lll-new}). 
Let $u = {n \choose k}$; we now define the r.v.s 
$\{C_{i,j}: i \in [m], j \in [u] \}$ as
follows. Fix any $i \in [m]$. Identify each $j \in [u]$ with some 
distinct $k$-element subset $S(j)$ of $[n]$, and let
\begin{equation}
\label{eqn:decomp}
C_{i,j} \doteq \frac{\prod_{v \in S(j)} Z_{i,v}}
{{{b_i + k} \choose k}}. 
\end{equation}
We now need to show that the r.v.s $C_{i,j}$
satisfy the conditions of Theorem~\ref{theorem:lll-new}. 
For any $i \in [m]$, let $\delta_i = k/b_i$.
Since $b_i \leq \yst$, we have, for 
each $i \in [m]$,
\begin{eqnarray*}
G(b_i, \delta_i) & \leq & G(\yst, k/\yst) \mbox{ (by Proposition
\ref{prop:G-simple})} \\ 
& \leq & G(\yst, H(\yst, 1/(et))) \\
& \leq & 1/(ekt) \mbox{ (by the definition of $H$).}
\end{eqnarray*}
Now by Theorem~\ref{theorem:sss}, we get

\begin{fact}
\label{fact1:Cij}
For all $i \in [m]$ and for all nonempty events $Z$,
$\Pr(E_i \bigm| Z) \leq \sum_{j \in [u]} \E[C_{i,j} \bigm| Z]$. Also,
$p_i \doteq \sum_{j \in [u]} \E[C_{i,j}] < G(b_i, \delta_i)
\leq 1/(ekt)$.
\end{fact}

Next since any $C_{i,j}$ involves (a product of) $k$ terms, each of
which ``depends'' on at most $(t-1)$ of the events 
$\{E_v: v \in ([m] - \{i\})\}$ by definition of $t$,
we see the important

\begin{fact}
\label{fact2:Cij}
$\forall  i \in [m] ~ \forall j \in [u]$, $C_{i,j} \in [0,1]$ and
$C_{i,j}$ ``depends'' on at most $d = k(t-1)$ of
the set of events $\{E_v: v \in ([m] - \{i\})\}$. 
\end{fact}
 From Facts~\ref{fact1:Cij} and \ref{fact2:Cij} and by noting that
$e p_i (d+1) \leq e (kt - k + 1)/(ekt) \leq 1$, we invoke 
Theorem~\ref{theorem:lll-new}, to see that
$\Pr(\bigwedge_{i \in [m]} \overline{E_i}) > 0$, concluding the proof
of Theorem~\ref{theorem:mip}. 
\end{proof}

Theorem~\ref{theorem:mip} gives good results if $t \ll m$, but can
we improve it further, say by replacing $t$ by $a$ ($\leq t$) in it?
As seen from (\ref{eqn:mip-ineq}), the key reason for 
$t \gg a^{\Theta(1)}$ is that $\mbox{max}_{i \in [n]}~ \ell_i \gg
a^{\Theta(1)}$. If we can essentially ``bring down'' 
$\mbox{max}_{i \in [n]}~ \ell_i$ by forcing many $x_{i,j}^*$ to be
zero for \textit{each} $i$, then we effectively 
reduce $t$ ($t \leq a \cdot max_i ~ \ell_i$, see (\ref{eqn:mip-ineq})); 
this is so since only
those $x_{i,j}^*$ that are neither zero nor one take part in the
rounding. A way of bootstrapping
Theorem~\ref{theorem:mip} to achieve this is shown by:

\begin{theorem}
\label{theorem:mip-bootstrap}
For any given MIP, there exists an integral solution of value at most
$\yst + O(1) + O(\min\{\yst,m\} \cdot H(\min\{\yst,m\},1/a))$. 
\end{theorem}

\begin{proof}
Let $K_0 > 0$ be a sufficiently large absolute constant. Now if
\begin{equation}
\label{eqn:boot-easy}
(\yst \geq t^{1/7}) \mbox{ or } (t \leq \max\{K_0, 2\}) 
\mbox{ or } (t \leq a^4)
\end{equation}
holds, then we will be done by Theorem~\ref{theorem:mip}. 
So we may assume that (\ref{eqn:boot-easy}) is
false. Also, if $\yst \leq
t^{-1/7}$, Theorem~\ref{theorem:mip} guarantees an integral solution
of value $O(1)$; thus, we also suppose that $\yst > t^{-1/7}$. 
The basic idea now is, as sketched
above, to set many $x_{i,j}^*$ to zero for each $i$ (without losing too
much on $\yst$), so that $\mbox{max}_i ~\ell_i$ and hence, $t$, will
essentially get reduced. Such an approach, whose performance will be validated
by arguments similar to those of Theorem~\ref{theorem:mip}, 
is repeatedly applied until (\ref{eqn:boot-easy}) holds, owing to the
(continually reduced) $t$ becoming small enough to satisfy 
(\ref{eqn:boot-easy}). There are two cases: \medskip

\noindent\textbf{Case I: $\yst \geq 1.$} Solve the LP relaxation, and set
$x_{i,j}' := (\yst)^2 (\log^5 t) x_{i,j}^*$. Conduct randomized rounding on the
$x_{i,j}'$ now, rounding \textit{each} $x_{i,j}'$ independently to 
$z_{i,j} \in \{\lfloor x_{i,j}' \rfloor, \lceil x_{i,j}' \rceil\}$. (Note the
key difference from Theorem~\ref{theorem:mip}, where for each $i$, we
round exactly one $x_{i,j}^*$ to $1$.) 

Let $K_1 > 0$ be a sufficiently large absolute constant.
We now use ideas similar to 
those used in our proof of 
Theorem~\ref{theorem:mip} to show that with nonzero probability, we have
both of the following:
\begin{eqnarray}
\forall i \in [m], & & 
(Az)_i \leq (\yst)^3 \log^5 t \cdot (1 + K_1 /((\yst)^{1.5} \log^2 t)), 
~~\textbf{and} \label{eqn:yst-around-mean} \\
~\forall i \in [n], & & 
| \sum_{j} z_{i,j} - (\yst)^2 \log^5 t | \leq K_1 \yst \log^3 t.
\label{sumj-zij-around-mean}
\end{eqnarray}
To show this, we proceed as follows. Let $E_1, E_2, \ldots, E_m$ be the
``bad'' events, one for each event in (\ref{eqn:yst-around-mean}) not holding; similarly, let
$E_{m+1}, E_{m+2}, \ldots, E_{m+n}$ be the ``bad'' events, one for each 
event in (\ref{sumj-zij-around-mean}) not holding. We want to use our extended LLL to show that with
positive probability, all these bad events can be avoided; specifically,
we need a way of decomposing each $E_i$ into a finite number of
non-negative r.v.s $C_{i,j}$. For each event
$E_{m+\ell}$ where $\ell \geq 1$, we define just one r.v.\ $C_{m+\ell, 1}$:
this is the indicator variable for the occurrence of $E_{m+\ell}$. 
For the events
$E_i$ where $i \leq m$, we decompose $E_i$ into r.v.s $C_{i,j}$ just as
in (\ref{eqn:decomp}): each such $C_{i,j}$ is now a scalar multiple of
at most
\[ O((\yst)^3 \log^5 t / ((\yst)^{1.5} \log^2 t)) =
O((\yst)^{1.5} \log^3 t) = O(t^{1.5/7} \log^3 t) \]
independent binary r.v.s that underlie our randomized rounding;
the second equality (big-Oh bound) here follows since (\ref{eqn:boot-easy})
has been assumed to not hold. Thus, it is easy to see that for all $i$, 
$1 \leq i \leq m + n$, and for
any $j$, the r.v.\ $C_{i,j}$ depends on at most 
\begin{equation}
\label{eqn:boot-dep}
O(t \cdot t^{1.5/7} \log^3 t)
\end{equation}
events $E_k$, where $k \not= i$. Also, as in our proof of
Theorem~\ref{theorem:mip}, Theorem~\ref{theorem:sss} gives a
direct proof of requirement (ii) of Theorem~\ref{theorem:lll-new};
part (b) of Theorem~\ref{theorem:sss} shows that for any 
desired constant $K$, we can choose
the constant $K_1$ large enough so that for all $i$,
$\sum_j \E[C_{i,j}] \leq t^{-K}$. Thus, in view of
(\ref{eqn:boot-dep}), we see by Theorem~\ref{theorem:lll-new} that
$\Pr(\bigwedge_{i=1}^{m+n} \overline{E_i}) > 0$.

Fix a rounding $z$ satisfying (\ref{eqn:yst-around-mean}) and 
(\ref{sumj-zij-around-mean}).
For each $i \in [n]$ and $j \in [\ell_i]$, we
renormalize as follows: $x_{i,j}'' := z_{i,j} / \sum_u z_{i,u}$. 
Thus we have $\sum_u x_{i,u}'' = 1$ for all $i$; we now see that we 
have two very
useful properties. First, since $\sum_j z_{i,j} \geq 
(\yst)^2 \log^5 t \cdot \left(1 - O(\frac{1}{\yst \log^2 t})\right)$ for all $i$
from (\ref{sumj-zij-around-mean}), we have, $\forall i \in [m]$, 
\begin{equation}
\label{eqn:yst-bound}
(Ax'')_i \leq \frac{\yst (1 + O(1/((\yst)^{1.5} \log^2 t)))}
{1 - O(1/(\yst \log^2 t))} \leq \yst(1 + O(1 /(\yst \log^2 t))).
\end{equation}
Second, since the $z_{i,j}$ are non-negative integers summing to
at most $(\yst)^2 \log^5 t (1 + O(1/(\yst \log^2 t)))$, \textit{at most
$O((\yst)^2 \log^5 t)$ values $x_{i,j}''$ are nonzero}, for each $i \in [n]$.
Thus, by losing a little in $\yst$ (see (\ref{eqn:yst-bound})), our 
``scaling up--rounding--scaling down'' method has given a fractional
solution $x''$ with a much-reduced $\ell_i$ for each $i$; $\ell_i$ is now
$O((\yst)^2 \log^5 t)$, essentially. Thus, $t$ 
has been reduced to $O(a (\yst)^2 \log^5 t)$; 
i.e., $t$ has been reduced to at most 
\begin{equation}
\label{eqn:new-t}
K_2 t^{1/4 + 2/7} \log^5 t
\end{equation}
for some constant $K_2 > 0$ that is independent of $K_0$, 
since (\ref{eqn:boot-easy}) was assumed false. Repeating this scheme
$O(\log \log t)$ times makes $t$ small enough to satisfy 
(\ref{eqn:boot-easy}). More formally, define $t_0 = t$, and 
$t_{i+1} = K_2 t_i^{1/4 + 2/7} \log^5 t_i$ for $i \geq 0$. Stop this
sequence at the first point where either $t = t_i$ satisfies
(\ref{eqn:boot-easy}), or $t_{i+1} \geq t_i$ holds. Thus, we finally
have $t$ small enough to satisfy (\ref{eqn:boot-easy}) or to be
bounded by some absolute constant. How much has $\max_{i \in [m]} (Ax)_i$ 
increased in the process? By (\ref{eqn:yst-bound}), we see that at the
end, 
\begin{equation}
\label{eqn:final-yst-bound}
\max_{i \in [m]} (Ax)_i 
\leq \yst \cdot \prod_{j \geq 0} (1 + O(1 /(y^* \log^2 t_j))) 
\leq \yst \cdot e^{O(\sum_{j \geq 0} 1 /(y^* \log^2 t_j))} \leq
\yst + O(1),
\end{equation}
since the values $\log t_j$ decrease geometrically and are lower-bounded
by some absolute positive constant. We may now apply 
Theorem~\ref{theorem:mip}. 
\medskip

\noindent\textbf{Case II: $t^{-1/7} < \yst < 1.$} The idea is the same
here, with the scaling up of $x_{i,j}^*$ 
being by $(\log^5 t) / \yst$; the same 
``scaling up--rounding--scaling down'' method works out. Since the ideas
are very similar to Case I, we only give a proof sketch here. 
We now scale up all the $x_{i,j}^*$ first by
$(\log^5 t) / \yst$ and do a randomized rounding. The analogs of
(\ref{eqn:yst-around-mean}) and (\ref{sumj-zij-around-mean}) that we now
want are
\begin{eqnarray}
\forall i \in [m], & & 
(Az)_i \leq \log^5 t \cdot (1 + K_1' /\log^2 t), 
~~\textbf{and} \label{case2:yst-around-mean} \\
~\forall i \in [n], & & 
| \sum_{j} z_{i,j} - \log^5 t / \yst | \leq K_1' \log^3 t /\sqrt{\yst}.
\label{case2:sumj-zij-around-mean}
\end{eqnarray}
Proceeding identically as in Case I, we can show that with positive
probability, (\ref{case2:yst-around-mean}) and
(\ref{case2:sumj-zij-around-mean}) hold simultaneously. Fix a rounding
where these two properties hold, and renormalize as before:
$x_{i,j}'' := z_{i,j} / \sum_u z_{i,u}$. Since 
(\ref{case2:yst-around-mean}) and
(\ref{case2:sumj-zij-around-mean}) hold, it is easy to show that the
following analogs of (\ref{eqn:yst-bound}) and (\ref{eqn:new-t}) hold:
\[ (Ax'')_i \leq \frac{\yst (1 + O(1/\log^2 t))}
{1 - O(\sqrt{y^*}/\log^2 t)} \leq \yst(1 + O(1 /\log^2 t)); ~~\mbox{and} \]
\[ \mbox{$t$ has been reduced to $O(a \log^5 t / \yst)$, i.e., to
$O(t^{1/4 + 1/7} \log^5 t)$.} \]
We thus only need $O(\log \log t)$ iterations, again. Also, the analog
of (\ref{eqn:final-yst-bound}) now is that
\[ \max_{i \in [m]} (Ax)_i 
\leq \yst \cdot \prod_{j \geq 0} (1 + O(1 /\log^2 t_j)) 
\leq \yst \cdot e^{O(\sum_{j \geq 0} 1 /\log^2 t_j)} \leq
\yst + O(1). \]

This completes the proof.
\end{proof}

We now study our improvements for discrepancy-type problems, which
are an important class of MIPs that, among other things, are useful
in devising divide-and-conquer algorithms. 
Given is a set-system $(X,F)$, where $X = [n]$ and
$F = \{D_1, D_2, \ldots, D_M\} \subseteq 2^X$.
Given a positive integer $\ell$, the problem is to partition 
$X$ into $\ell$ parts, so that \textit{each} $D_j$ is ``split well'':
we want a function $f: X \rightarrow [\ell]$ which minimizes
$\mbox{max}_{j \in [M], k \in [\ell]} ~ |\{i \in D_j: ~f(i) = k\}|$.
(The case $\ell = 2$ is the standard set-discrepancy
problem.) To motivate this problem, suppose we have a (di)graph $(V,A)$;
we want a partition of $V$ into $V_1, \ldots, V_{\ell}$ such that
$\forall v \in V$, 
$\{|\{j \in N(v) \cap V_k\}|:~ k \in [\ell]\}$
are ``roughly the same'', where $N(v)$ is the (out-)neighborhood of
$v$. See, e.g., \cite{alon:strong-chi,karloff-shmoys} for how
this helps construct divide-and-conquer approaches. 
This problem is naturally modeled by the above set-system problem.

Let $\Delta$ be the degree of $(X,F)$, i.e., $\mbox{max}_{i \in [n]}
|\{j: i \in D_j\}|$, and let $\Delta' \doteq 
\mbox{max}_{D_j \in F} ~ |D_j|$. Our problem is naturally
written as an MIP with $m = M\ell$, $\ell_i = \ell$ for each $i$, and
$g = a = \Delta$, in the notation of
Definition~\ref{defn:mip}; $\yst = \Delta'/\ell$ here.
The analysis of \cite{raghavan-thompson}
gives an integral solution of value at most 
$\yst(1 + O(H(\yst, 1/(M\ell))))$, while \cite{klrtvv} presents a
solution of value at most $\yst + \Delta$. Also, since any
$D_j \in F$ intersects at most $(\Delta - 1)\Delta'$ other elements of
$F$, Lemma~\ref{lemma:lll} shows that randomized rounding
produces, with positive probability, a solution of value at most
$\yst(1 + O(H(\yst, 1/(e \Delta' \Delta \ell))))$. 
This is the approach taken
by \cite{furedi-kahn} for their case of interest: $\Delta = \Delta'$,
$\ell = \Delta / \log \Delta$. 

Theorem~\ref{theorem:mip-bootstrap} shows the existence of an integral
solution of value
$\yst(1 + O(H(\yst, 1/\Delta))) + O(1)$, i.e., 
\textit{removes the dependence on
$\Delta'$}. This is an improvement on all the three results above.
As a specific interesting case, suppose $\ell$ grows at most as fast as 
$\Delta' / \log \Delta$. Then we see that good integral solutions--those 
that grow at the rate of $O(\yst)$ or better--exist, 
and this was not known before. (The approach of \cite{furedi-kahn}
shows such a result for 
$\ell = O(\Delta' / \log(\mbox{max}\{\Delta,\Delta'\}))$. Our bound of
$O(\Delta' / \log \Delta)$ is always better than this, and especially so if
$\Delta' \gg \Delta$.)

\section{Approximating Covering Integer Programs}
\label{section:cip}
One of the main ideas behind Theorem~\ref{theorem:lll-new} was to
extend the basic inductive proof behind the LLL by 
decomposing the ``bad'' events $E_i$ appropriately into the r.v.s
$C_{i,j}$. We now use this general idea in a different context,
that of (multi-criteria) covering integer programs, with an additional
crucial ingredient being a useful correlation inequality, the FKG
inequality \cite{fkg}. The reader is asked to recall the discussion of
(multi-criteria) CIPs from \S~\ref{section:ip-defs}. 
We start with a discussion of randomized rounding for CIPs, the
Chernoff lower-tail bound, and the FKG inequality in 
\S~\ref{sec:prelims}. These lead to our improved, but nonconstructive,
approximation bound for column-sparse (multi-criteria) CIPs, in
\S~\ref{section:nonconstr}. This is then made constructive in
\S~\ref{section:construct}; we also discuss there what we view
as novel about this constructive approach. 

\subsection{Preliminaries}
\label{sec:prelims}
Let us start with a simple and well-known approach to tail bounds.
Suppose $Y$ is a random variable and $y$ is some value. Then, for any
$0 \leq \delta < 1$, we have
\begin{equation}
\label{eqn:lower-tail-gen}
\Pr(Y \leq y) \leq
\Pr((1 - \delta)^Y \geq (1 - \delta)^{y}) 
\leq 
\frac{\E[(1 - \delta)^Y]}{(1 - \delta)^{y}},
\end{equation}
where the inequality is a consequence of Markov's inequality. 

We next setup some basic notions related to approximation algorithms
for (multi-criteria) CIPs. Recall that in such problems,
we have $\ell$ given non-negative
vectors $c_1, c_2, \ldots, c_{\ell}$ such that
for all $i$, $c_i \in [0,1]^n$ with max$_j~ c_{i,j} = 1$;
$\ell = 1$ in the case of CIPs. 
Let $x = (x_1^*, x_2^*, \ldots, x_n^*)$ denote a given fractional
solution that satisfies the system of constraints $Ax \geq b$. 
We are not concerned here with
how $x^*$ was found: typically, $x^*$ would be an optimal solution
to the LP relaxation of the problem. (The LP relaxation is obvious if,
e.g., $\ell = 1$, or, say, if the given
multi-criteria aims to minimize $\max_i c_i^T \cdot x^*$, or to keep
each $c_i^T \cdot x^*$ bounded by some target value $v_i$.)
We now consider how to
round $x^*$ to some integral $z$ so that: 
\begin{description}
\item[\textbf(P1)] the constraints $Az \geq b$ hold, and 
\item[\textbf(P2)] for all $i$, $c_i^T \cdot z$ is 
``not much bigger'' than $c_i^T \cdot x^*$: our approximation bound
will be a measure of how small a ``not much bigger value'' we can achieve
in this sense. 
\end{description}

Let us now discuss the ``standard'' randomized rounding scheme for
(multi-criteria) CIPs. 
We assume a fixed instance as well as $x^*$, from now on.
For an $\alpha > 1$
to be chosen suitably, set $x_j' = \alpha x_j^*$, for each $j \in [n]$.
We then construct a random integral solution $z$ by setting, 
independently for each $j \in [n]$,
$z_j = \lfloor x_j' \rfloor + 1$ with probability 
$x_j' - \lfloor x_j' \rfloor$, and $z_j = \lfloor x_j' \rfloor$
with probability $1 - (x_j' - \lfloor x_j' \rfloor)$. 
The aim then is to show that with positive (hopefully high) probability,
\textbf{(P1)} and \textbf{(P2)} happen simultaneously. We now introduce
some useful notation. 
For every $j \in [n]$, let $s_j = \lfloor x_j' \rfloor$.
Let $A_i$ denote the $i$th row of $A$, and 
let $X_1, X_2, \ldots, X_n \in \zo$ be \textit{independent} r.v.s
with $\Pr(X_j = 1) = x_j' - s_j$ for all $j$.
The bad event $E_i$ that the $i$th constraint is violated by our
randomized rounding is given by
$E_i \equiv ``A_i \cdot X < \mu_i(1 - \delta_i)$'',
where $\mu_i = \E[A_i \cdot X]$ and 
$\delta_i = 1 - (b_i - A_i \cdot s)/\mu_i$. 
We now bound $\Pr(E_i)$ for all $i$, when the standard randomized 
rounding is used.

\begin{lemma}
\label{lemma:pmax}
Define $g(B,\alpha) \doteq (\alpha \cdot e^{-(\alpha - 1)})^B$.
Then for all $i$, 
\[ \Pr(E_i) \leq \frac{\E[(1 - \delta_i)^{A_i \cdot X}]}
{(1 - \delta_i)^{(1 - \delta_i)\mu_i}}
\leq g(B,\alpha) \leq e^{-B (\alpha - 1)^2 / (2\alpha)} \]
under standard randomized rounding.
\end{lemma}

\begin{proof}
The first inequality follows from
(\ref{eqn:lower-tail-gen}). Next, the 
Chernoff-Hoeffding lower-tail approach \cite{alon-spencer,motwani-raghavan}
shows that 
\[ \frac{\E[(1 - \delta_i)^{A_i \cdot X}]}
{(1 - \delta_i)^{(1 - \delta_i)\mu_i}}
\leq 
\left(\frac{e^{-\delta_i}}{(1 - \delta_i)^{1 - \delta_i}}\right)^{\mu_i}. \]
It is observed in \cite{srin:pos-correl} (and is not hard to see) that
this latter quantity is
maximized when $s_j = 0$ for all $j$, and when each $b_i$ equals
its minimum value of $B$. Thus we see that
$\Pr(E_i) \leq g(B,\alpha)$. The inequality 
$g(B,\alpha) \leq e^{-B (\alpha - 1)^2 / (2\alpha)}$ for $\alpha \geq 1$,
is well-known and easy to verify via elementary calculus. 
\end{proof}

Next, the FKG inequality is a useful correlation inequality, 
a special case of which is as follows \cite{fkg}. Given binary vectors
$\vec{a} = (a_1, a_2, \ldots, a_{\ell}) \in \{0,1\}^{\ell}$ and
$\vec{b} = (b_1, b_2, \ldots, b_{\ell}) \in \{0,1\}^{\ell}$, let us
partially order them by coordinate-wise domination:
$\vec{a} \preceq \vec{b}$ iff $a_i \leq b_i$ for all $i$. 
Now suppose $Y_1, Y_2, \ldots, Y_{\ell}$ are \textit{independent} r.v.s, 
each taking values in $\{0,1\}$. Let $\vec{Y}$ denote the
vector $(Y_1, Y_2, \ldots, Y_{\ell})$. Suppose an event 
$\mathcal{A}$ is completely defined by the value of
$\vec{Y}$. Define $\mathcal{A}$ to be \textit{increasing} iff: for 
all $\vec{a} \in
\{0,1\}^{\ell}$ such that $\mathcal{A}$ holds 
when $\vec{Y} = \vec{a}$, $\mathcal{A}$
also holds when $\vec{Y} = \vec{b}$, for any $\vec{b}$ such that 
$\vec{a} \preceq \vec{b}$. Analogously,
event $\mathcal{A}$ is \textit{decreasing} iff: for all $\vec{a} \in
\{0,1\}^{\ell}$ such that $\mathcal{A}$ holds 
when $\vec{Y} = \vec{a}$, $\mathcal{A}$
also holds when $\vec{Y} = \vec{b}$, for any $\vec{b} \preceq \vec{a}$. 
The FKG inequality proves certain intuitively appealing bounds:

\begin{lemma}
\label{lemma:fkg}
\textbf{(FKG inequality)} Let $I_1, I_2, \ldots, I_t$ be any sequence 
of \textit{increasing} events and $D_1, D_2, \ldots, D_t$ be any sequence
of \textit{decreasing} events (each $I_i$ and $D_i$ completely determined
by $\vec{Y}$). Then for any $i \in [t]$ and any $S \subseteq [t]$,

\noindent (i) $\Pr(I_i | \bigwedge_{j \in S} I_j) \geq \Pr(I_i)$ and
$\Pr(D_i | \bigwedge_{j \in S} D_j) \geq \Pr(D_i)$;

\noindent (ii) $\Pr(I_i | \bigwedge_{j \in S} D_j) \leq \Pr(I_i)$ and
$\Pr(D_i | \bigwedge_{j \in S} I_j) \leq \Pr(D_i)$.
\end{lemma} 

Returning to our random variables $X_j$ and events $E_i$, we get
the following lemma as an easy consequence of the FKG inequality,
since each event of the form ``$\overline{E_i}$'' or
``$X_j = 1$'' is an increasing event as a function
of the vector $(X_1, X_2, \ldots, X_n)$: 

\begin{lemma}
\label{lemma:correl}
For all $B_1, B_2 \subseteq [m]$ such that
$B_1 \cap B_2 = \emptyset$ and for any $B_3 \subseteq [n]$,
$\Pr(\bigwedge_{i \in B_1} \overline{E_i} \bigm|
((\bigwedge_{j \in B_2} \overline{E_j}) \wedge
(\bigwedge_{k \in B_3} (X_k = 1))) 
 \geq 
\prod_{i \in B_1}\Pr(\overline{E_i})$.
\end{lemma}

\subsection{Nonconstructive approximation bounds for (multi-criteria) CIPs}
\label{section:nonconstr}
\begin{definition}
\label{defn:calR}
\textbf{(The function $\mathcal{R}$)}
For any $s$ and any $j_1 < j_2 < \cdots < j_{s}$, let
$\mathcal{R}(j_1, j_2, \ldots, j_s)$ be the set of indices
$i$ such that row $i$ of the constraint system ``$Ax \geq b$''
has at least one of the variables $j_k, ~1 \leq k \leq s$,
appearing with a nonzero coefficient. (Note from 
the definition of $a$ in Defn.~\ref{defn:cip}, that
$|\mathcal{R}(j_1, j_2, \ldots, j_s)| \leq a \cdot s$.)
\end{definition}

Let the vector $x^* = (x_1^*, x_2^*, \ldots, x_n^*)$, the parameter
$\alpha > 1$, and the ``standard'' randomized rounding 
scheme, be as defined in \S~\ref{sec:prelims}. The standard rounding
scheme is sufficient for our (nonconstructive) purposes now; 
we generalize this scheme as follows, for later use 
in \S~\ref{section:construct}. 

\begin{definition}
\textbf{(General randomized rounding)}
Given a vector $p = (p_1, p_2, \ldots, p_n) \in [0,1]^n$,
the \textit{general randomized rounding with parameter $p$} generates
\textit{independent} random variables 
$X_1, X_2, \ldots, X_n \in \zo$ with $\Pr(X_j = 1) = p_j$; the
rounded vector $z$ is defined by $z_j = 
\lfloor \alpha x_j^* \rfloor + X_j$ for all $j$.
(As in the standard rounding, we set each $z_j$ to be either
$\lfloor \alpha x_j^* \rfloor$ or $\lceil \alpha x_j^* \rceil$;
the standard rounding is the special case 
in which $\E[z_j] = \alpha x_j^*$ for all $j$.)
\end{definition}

We now present an important lemma, 
Lemma~\ref{lemma:anti-fkg}, to get correlation inequalities which
``point'' in the ``direction'' opposite to FKG.
Some ideas from the proof of Lemma~\ref{lemma:lll} 
will play a crucial role in our proof this lemma. 

\begin{lemma}
\label{lemma:anti-fkg}
Suppose we employ general randomized rounding 
with some parameter $p$, and that
$\Pr(\bigwedge_{i=1}^m \overline{E_i})$ 
is nonzero under this rounding. 
The following hold for any $q$ and any 
$1 \leq j_1 < j_2 < \cdots < j_{q} \leq n$.

(i) 
\begin{equation}
\label{anti-fkg:abstract}
\Pr(X_{j_1} = X_{j_2} = \cdots = X_{j_{q}} = 1 \bigm| 
\bigwedge_{i=1}^m \overline{E_i}) 
\leq  
\frac{\prod_{t=1}^q p_{j_t}}
{\prod_{i \in \mathcal{R}(j_1, j_2, \ldots, j_q)} (1 - \Pr(E_i))};
\end{equation}
the events $E_i \equiv ((Az)_i < b_i)$ are defined here w.r.t.\ the
general randomized rounding. 

(ii) In the special case of standard randomized rounding,
\begin{equation}
\label{anti-fkg:concrete}
\prod_{i \in \mathcal{R}(j_1, j_2, \ldots, j_q)} (1 - \Pr(E_i))
\geq (1 - g(B,\alpha))^{aq};
\end{equation}
the function $g$ is as defined in Lemma~\ref{lemma:pmax}. 
\end{lemma}

\vspace*{-0.03in}
\begin{proof} 
(i) Note first that if we wanted a \textit{lower} bound on the l.h.s.,
the FKG inequality would immediately imply that the l.h.s.\ is
at least $p_{j_1} p_{j_2} \cdots p_{j_q}$. We get around this
``correlation problem'' as follows. 
Let $Q = \mathcal{R}(j_1, j_2, \ldots, j_q)$; 
let $Q' = [m] - Q$. Let 
\[ Z_1 \equiv (\bigwedge_{i \in Q} \overline{E_i}), ~\mbox{and}~
Z_2 \equiv (\bigwedge_{i \in Q'} \overline{E_i}). \]
Letting $Y = \prod_{t=1}^q X_{j_t}$, note that
\begin{equation}
\label{eqn:cip1}
|Q| \leq aq ~\mbox{ and}
\end{equation}
\begin{equation}
\label{eqn:cip2}
Y \mbox{ is independent of } Z_2. 
\end{equation}
Now, 
\begin{eqnarray*}
\Pr(Y = 1\bigm| (Z_1 \wedge Z_2)) & = & 
\frac{\Pr(((Y = 1) \wedge Z_1) \bigm| Z_2)}{\Pr(Z_1 \bigm| Z_2)} \\
& \leq & \frac{\Pr((Y = 1) \bigm| Z_2)}{\Pr(Z_1 \bigm| Z_2)} \\
& = & \frac{\Pr(Y = 1)}{\Pr(Z_1 \bigm| Z_2)} 
~~\mbox{(by (\ref{eqn:cip2}))} \\
& \leq & 
\frac{\prod_{t=1}^q \Pr(X_{j_t} = 1)}
{\prod_{i \in \mathcal{R}(j_1, j_2, \ldots, j_q)} (1 - \Pr(E_i))} 
~~\mbox{(by Lemma~\ref{lemma:correl})}.
\end{eqnarray*}

(ii) We get (\ref{anti-fkg:concrete}) from Lemma~\ref{lemma:pmax} and
(\ref{eqn:cip1}). 
\end{proof}

We will use Lemmas~\ref{lemma:correl} and \ref{lemma:anti-fkg} 
to prove Theorem~\ref{theorem:mult-cip}. As a warmup, let us
start with a result for the special case of CIPs; recall that
$y^*$ denotes $c^T \cdot x^*$. 

\begin{theorem}
\label{theorem:cip}
For any given CIP, suppose we choose $\alpha, \beta > 1$ such that
$\beta(1 - g(B,\alpha))^a > 1$. Then, there exists a 
feasible solution of value at most $\yst \alpha\beta$. 
In particular, there is an absolute constant $K > 0$ such that if
$\alpha,\beta > 1$ are chosen as: 
\begin{eqnarray}
\alpha & = & K \cdot \ln (a+1) / B \mbox{ and } \beta = 2, 
\mbox{ if } \ln (a+1) \geq B, \mbox{ and } \label{cip:case1} \\
\alpha & = & \beta = 1 + K \cdot \sqrt{\ln (a+1)/B}, 
\mbox{ if } \ln (a+1) < B; \label{cip:case2}
\end{eqnarray}
then, there exists a feasible solution of value at most $\yst \alpha\beta$.
Thus, the integrality gap is at most 
$1 + O(\mbox{max}\{\ln (a+1)/B, \sqrt{\ln (a+1)/B}\})$.
\end{theorem}

\begin{proof}
Conduct standard randomized rounding, and let $\mathcal{E}$ be
the event that $c^T \cdot z > y^* \alpha \beta$. 
Setting $Z \equiv \bigwedge_{i \in [m]} \overline{E_i}$ and 
$\mu \doteq \E[c^T \cdot z] = \yst\alpha$, we see by Markov's inequality
that $\Pr(\mathcal{E} \bigm| Z)$ is at most
$R = (\sum_{j=1}^{n} c_j \Pr(X_j = 1 \bigm| Z))/(\mu\beta)$.
Note that $\Pr(Z) > 0$ since $\alpha > 1$; so, we now
seek to make $R < 1$, which will complete the proof. 
Lemma~\ref{lemma:anti-fkg} shows that 
\[ R \leq \frac{\sum_j c_j p_j}{\mu\beta \cdot (1 - g(B,\alpha))^a}
= \frac{1}{\beta(1 - g(B,\alpha))^a}; \]
thus, the condition 
$\beta(1 - g(B,\alpha))^a > 1$ suffices. 

Simple algebra shows that choosing $\alpha, \beta
> 1$ as in (\ref{cip:case1}) and (\ref{cip:case2}), 
ensures that $\beta(1 - g(B,\alpha))^a > 1$.
\end{proof}

The basic approach of our proof of Theorem~\ref{theorem:cip} is to
follow the main idea of Theorem~\ref{theorem:lll-new}, and
to decompose the event ``$\mathcal{E} \bigm| Z$'' into a non-negative
linear combination of events of the form ``$X_j = 1 \bigm| Z$'; we then
exploited the fact that each $X_j$ depends on at most $a$ of the events
comprising $Z$. We now extend Theorem~\ref{theorem:cip} and also 
generalize to multi-criteria CIPs. Instead of employing
just a ``first moment method'' (Markov's inequality) as in the
proof of Theorem~\ref{theorem:cip}, we will work with 
higher moments: the functions $S_k$ defined in
(\ref{eqn:S_j}) and used in Theorem~\ref{theorem:sss}.
Suppose some parameters $\lambda_i > 0$ are
given, and that our goal is to round $x^*$ to $z$ so that the event
\begin{equation}
\label{eqn:A-def}
\mathcal{A} \equiv ``(Az \geq b) ~\wedge~
(\forall i, ~c_i^T \cdot z \leq \lambda_i)''
\end{equation}
holds. We first give a sufficient condition for this to hold, in
Theorem~\ref{theorem:mult-cip}; we then derive some concrete
consequences in Corollary~\ref{cor:mult-cip}.
We need one further definition before presenting
Theorem~\ref{theorem:mult-cip}. Recall that $A_i$ and $b_i$
respectively denote the $i$th row of $A$ and the $i$th component
of $b$. Also, the vector $s$ and values $\delta_i$ will throughout 
be as in the definition of \textbf{standard} randomized rounding. 

\begin{definition}
\label{defn:ch}
\textbf{(The functions $\CH$ and $\CH'$)} 
Suppose we conduct general randomized rounding 
with some parameter $p$; i.e., let $X_1, X_2, \ldots, X_n$ be
independent binary random variables such that
$\Pr(X_j = 1) = p_j$. For each $i \in [m]$, define
\[ \CH_i(p) \doteq 
\frac{\E[(1 - \delta_i)^{A_i \cdot X}]}
{(1 - \delta_i)^{b_i - A_i \cdot s}} 
= \frac{\prod_{j \in [n]} \E[(1 - \delta_i)^{A_{i,j} X_j}]}
{(1 - \delta_i)^{b_i - A_i \cdot s}} 
~~\mbox{and}~~
\CH'_i(p) \doteq \min\{CH_i(p), 1\}. \]
(Note from (\ref{eqn:lower-tail-gen}) that if we 
conduct general randomized rounding with parameter $p$,
then 
$\Pr((Az)_i < b_i) \leq \CH_i(p) \leq \CH'_i(p)$; also,
``$\CH$'' stands for ``Chernoff-Hoeffding''.)
\end{definition}

\begin{theorem}
\label{theorem:mult-cip}
Suppose we are given a multi-criteria CIP, as well as some
parameters $\lambda_1, \lambda_2, \ldots, \lambda_{\ell} > 0$.
Let $\mathcal{A}$ be as in (\ref{eqn:A-def}). Then, for any
sequence of positive integers $(k_1, k_2, \ldots, k_{\ell})$
such that $k_i \leq \lambda_i$, the following
hold.

\smallskip \noindent (i) Suppose we employ general randomized rounding 
with parameter $p = (p_1, p_2, \ldots, p_n)$. 
Then, $\Pr(\mathcal{A})$ is at least
\begin{equation}
\label{phi1}
\Phi(p) \doteq 
\left(\prod_{r \in [m]} (1 - \CH'_r(p))\right) - 
\sum_{i=1}^{\ell} \frac{1}{{{\lambda_i} \choose k_i}} \cdot
\sum_{j_1 < \cdots < j_{k_i}} 
\left(\prod_{t=1}^{k_i} c_{i,j_t} \cdot p_{j_t} \right) \cdot
\prod_{r \not\in \mathcal{R}(j_1, \ldots, j_{k_i})}
(1 - \CH'_r(p)); 
\end{equation}
as in Defn.~\ref{defn:multi-cip}, $c_{i,j} \in [0,1]$ is the $j$th
coordinate of the vector $c_i$. 

\smallskip \noindent (ii) Suppose we employ the standard randomized 
rounding to get a rounded vector $z$.
Let $\lambda_i = \nu_i(1 + \gamma_i)$ for each $i \in [\ell]$, where
$\nu_i = \E[c_i^T \cdot z] = \alpha \cdot (c_i^T \cdot x^*)$ and
$\gamma_i > 0$ is some parameter. 
Then, 
\begin{equation}
\Phi(p) \geq (1 - g(B,\alpha))^m \cdot
\left(1 - \sum_{i=1}^{\ell} \frac{{n \choose {k_i}} \cdot (\nu_i/n)^{k_i}}
{{{\nu_i(1 + \gamma_i)} \choose k_i}} \cdot
(1 - g(B,\alpha))^{- a \cdot k_i}\right). 
\label{eqn:std-phi}
\end{equation}
In particular, if the r.h.s.\ of (\ref{eqn:std-phi}) is positive, then
$\Pr(\mathcal{A}) > 0$ for standard randomized rounding. 
\end{theorem}

The proof is a simple generalization of that of
Theorem~\ref{theorem:cip}, and is 
deferred to Section~\ref{sec:deferred-proofs}.
Theorem~\ref{theorem:cip} is the special case of
Theorem~\ref{theorem:mult-cip} corresponding to $\ell = k_1 = 1$.
To make the general result of Theorem~\ref{theorem:mult-cip} 
more concrete, we now study an additional special case. We present
this special case as one possible ``proof of concept'', rather than
as an optimized one; e.g., the constant ``$3$'' in the bound
``$c_i^T \cdot z \leq 3\nu_i$'' can be improved.

\begin{corol}
\label{cor:mult-cip}
There is an absolute constant $K' > 0$ such that the following
holds. Suppose we are given a multi-criteria CIP with notation as in
part (ii) of Theorem~\ref{theorem:mult-cip}. Define 
$\alpha = K' \cdot \max\{\frac{\ln(a) + \ln\ln(2\ell)}{B}, ~1\}$.
Now if $\nu_i \geq \log^2 (2 \ell)$ for all $i \in [\ell]$,
then standard randomized rounding produces a feasible solution $z$
such that $c_i^T \cdot z \leq 3\nu_i$ for all $i$, with 
positive probability.

In particular, this can be shown by setting 
$k_i = \lceil \ln(2 \ell) \rceil$ and $\gamma_i = 2$ for all $i$,
in part (ii) of Theorem~\ref{theorem:mult-cip}. 
\end{corol}

\begin{proof}
Let us employ Theorem~\ref{theorem:mult-cip}(ii) with
$k_i = \lceil \ln(2 \ell) \rceil$ and $\gamma_i = 2$ for all $i$.
We just need to establish that the r.h.s.\ of (\ref{eqn:std-phi}) 
is positive. We need to show that
\[ \sum_{i=1}^{\ell} \frac{{n \choose {k_i}} \cdot (\nu_i/n)^{k_i}}
{{{3 \nu_i} \choose k_i}} \cdot
(1 - g(B,\alpha))^{- a \cdot k_i} < 1; \]
it is sufficient to prove that for all $i$,
\begin{equation}
\label{eqn:cov-cor}
\frac{\nu_i^{k_i} / k_i!}{{{3 \nu_i} \choose k_i}} \cdot
(1 - g(B,\alpha))^{- a \cdot k_i} < 1/\ell. 
\end{equation}
We make two observations now.
\begin{itemize}
\item Since $k_i \sim \ln \ell$ and $\nu_i \geq \log^2 (2 \ell)$,
\[ {{3 \nu_i} \choose k_i} = (1/k_i!) \cdot
\prod_{j=0}^{k_i - 1} (3 \nu_i - j) = 
(1/k_i!) \cdot (3 \nu_i)^{k_i} \cdot 
e^{-\Theta(\sum_{j=0}^{k_i - 1} j/\nu_i)} = 
\Theta((1/k_i!) \cdot (3 \nu_i)^{k_i}). \]
\item $(1 - g(B,\alpha))^{- a \cdot k_i}$ can be made arbitrarily
close to $1$ by choosing the constant $K'$ large enough.
\end{itemize}
These two observations establish (\ref{eqn:cov-cor}). 
\end{proof}

\subsection{Constructive version}
\label{section:construct}
It can be shown that for many problems,
randomized rounding produces the solutions shown to exist by
Theorem~\ref{theorem:cip} and Theorem~\ref{theorem:mult-cip}, with very 
low probability: e.g., probability almost exponentially small in the
input size. Thus we need to obtain constructive versions of 
these theorems. Our method will be a deterministic procedure that
makes $O(n)$ calls to the function $\Phi(\cdot)$, in addition to
$\poly(n,m)$ work. Now, if $k'$ denotes the maximum of all the
$k_i$, we see that $\Phi$ can be evaluated in
$\poly(n^{k'},m)$ time. Thus, our overall procedure runs in time
$\poly(n^{k'},m)$ time. In particular, we get constructive versions
of Theorem~\ref{theorem:cip} and Corollary~\ref{cor:mult-cip} that
run in time $\poly(n,m)$ and $\poly(n^{\log \ell},m)$, respectively. 

Our approach is as follows. We start with a vector $p$ that
corresponds to standard randomized rounding, for which we know
(say, as argued in Corollary~\ref{cor:mult-cip}) that $\Phi(p) > 0$.
In general, we have a vector of probabilities 
$p = (p_1, p_2, \ldots, p_n)$ such that $\Phi(p) > 0$. If
$p \in \{0,1\}^n$, we are done. Otherwise suppose some $p_j$ lies
in $(0,1)$; by renaming the variables,
we will assume without loss of generality that $j=n$. 
Define $p' = (p_1, p_2, \ldots, p_{n-1}, 0)$ and
$p'' = (p_1, p_2, \ldots, p_{n-1}, 1)$. The main fact we wish to show
is that $\Phi(p') > 0$ or $\Phi(p'') > 0$: 
we can then set $p_n$ to $0$ or $1$ 
appropriately, and continue. 
(As mentioned in the previous paragraph, we thus have
$O(n)$ calls to the function $\Phi(\cdot)$ in total.)
Note that although some of the $p_j$ will
lie in $\{0,1\}$, we can crucially continue to view the $X_j$ as
\textit{independent} random variables with $\Pr(X_j = 1) = p_j$. 

So, our main goal is: assuming that $p_n \in (0,1)$ and that
\begin{equation}
\label{constr:ind-assumption}
\Phi(p) > 0, 
\end{equation}
to show that $\Phi(p') > 0$ or $\Phi(p'') > 0$. In order
to do so, we make some observations and introduce some simplifying
notation. Define, for each $i \in [m]$: 
$q_i = \CH'_i(p)$, 
$q_i' = \CH'_i(p')$, and
$q_i'' = \CH'_i(p'')$.
Also define the vectors $q \doteq (q_1, q_2, \ldots, q_m)$,
$q' \doteq (q_1', q_2', \ldots, q_m')$, and
$q'' \doteq (q_1'', q_2'', \ldots, q_m'')$. 
We now present a useful lemma about these vectors:

\begin{lemma}
\label{lemma:q-vectors}
For all $i \in [m]$, we have
\begin{eqnarray}
0 & \leq & q_{i}'' \leq q_{i}' \leq 1; \label{eqn:qi1-small} \\
q_i & \geq & p_n q_{i}'' + (1-p_n) q_i'; ~\mbox{and}
\label{eqn:q-conv-comb} \\
q_i' & = & q_{i}'' = q_i ~\mbox{if}~ i \not\in \mathcal{R}(n). 
\label{eqn:q-inotinRn}
\end{eqnarray}
\end{lemma}

\begin{proof}
The proofs of (\ref{eqn:qi1-small}) and 
(\ref{eqn:q-inotinRn}) are straightforward. As for
(\ref{eqn:q-conv-comb}), we proceed as in \cite{srin:pos-correl}.
First of all, if $q_i = 1$, then we are done, since
$q_{i}'', ~q_i' \leq 1$. So suppose $q_i < 1$; in this case,
$q_i = \CH_i(p)$. Now, Definition~\ref{defn:ch} shows that
\[ \CH_i(p) = p_n \CH_i(p'') + (1-p_n) \CH_i(p'). \]
Therefore,
$q_i = \CH_i(p) = p_n \CH_i(p'') + (1-p_n) \CH_i(p') \geq
p_n \CH'_i(p'') + (1-p_n) \CH'_i(p')$. 
\end{proof}

Since we are mainly concerned with the vectors $p$, $p'$ and $p''$
now, we will view the values $p_1, p_2, \ldots, p_{n-1}$ as
arbitrary but \textit{fixed}, subject to (\ref{constr:ind-assumption}). 
The function $\Phi(\cdot)$ now has a simple form; to see this, we first
define, for a vector $r = (r_1, r_2, \ldots, r_m)$ and a
set $U \subseteq [m]$,
\[ f(U, r) = \prod_{i \in U} (1 - r_i). \]
Recall that $p_1, p_2, \ldots, p_{n-1}$ are considered as constants now.
Then, it is evident from (\ref{phi1}) that there exist constants
$u_1, u_2, \ldots, u_t$ and $v_1, v_2, \ldots, v_{t'}$, as well as
subsets $U_1, U_2, \ldots, U_t$ and $V_1, V_2, \ldots, V_{t'}$
of $[m]$, such that
\begin{eqnarray}
\Phi(p) & = & f([m], q) - (\sum_i u_i \cdot f(U_i, q)) 
- (p_n \cdot \sum_j v_j \cdot f(V_j, q)); \label{phip} \\
\Phi(p') & = & f([m], q') - (\sum_i u_i \cdot f(U_i, q')) 
- (0 \cdot \sum_j v_j \cdot f(V_j, q'))
= f([m], q') - \sum_i u_i \cdot f(U_i, q'); \label{phip'} \\
\Phi(p'') & = & f([m], q'') - (\sum_i u_i \cdot f(U_i, q'')) 
- (1 \cdot \sum_j v_j \cdot f(V_j, q'')) \nonumber \\
& = & f([m], q'') - (\sum_i u_i \cdot f(U_i, q'')) 
- (\sum_j v_j \cdot f(V_j, q'')). 
\label{phip''}
\end{eqnarray}
Importantly, we also have the following:
\begin{equation}
\label{eqn:uvV}
\mbox{the constants $u_i, v_j$ are non-negative}; ~~
\forall j, ~V_j \cap \mathcal{R}(n) = \emptyset.
\end{equation}

Recall that our goal is to show that
$\Phi(p') > 0$ or $\Phi(p'') > 0$. 
We will do so by proving that 
\begin{equation}
\label{eqn:goal-conv-comb}
\Phi(p) \leq p_n \Phi(p'') + (1-p_n) \Phi(p').
\end{equation}
Let us use the equalities 
(\ref{phip}), (\ref{phip'}), and (\ref{phip''}).
In view of (\ref{eqn:q-inotinRn}) and
(\ref{eqn:uvV}), the term
``$- p_n \cdot \sum_j v_j \cdot f(V_j, q)$'' 
on both sides of the inequality 
(\ref{eqn:goal-conv-comb}) cancels; defining
$\Delta(U) \doteq 
(1 - p_n) \cdot f(U, q') + p_n \cdot f(U, q'') - f(U, q)$,
inequality (\ref{eqn:goal-conv-comb}) reduces to
\begin{equation}
\label{eqn:aim2}
\Delta([m]) - \sum_{i} u_i \cdot \Delta(U_i) \geq 0.
\end{equation}

Before proving this, we pause to note a challenge we face. Suppose
we only had to show that, say, $\Delta([m])$ is non-negative; this
is exactly the issue faced in \cite{srin:pos-correl}. Then, we will
immediately be done by part (i) of Lemma~\ref{lemma:Delta}, which
states that $\Delta(U) \geq 0$ for any set $U$. However, 
(\ref{eqn:aim2}) also has terms such as ``$u_i \cdot \Delta(U_i)$''
with a \textit{negative} sign in front. To deal with this, we need 
something more than just that $\Delta(U) \geq 0$ for all $U$; we
handle this by part (ii) of Lemma~\ref{lemma:Delta}. We view this
as the main novelty in our constructive version here. 

\begin{lemma}
\label{lemma:Delta}
Suppose $U \subseteq V \subseteq [m]$. Then,
(i) $\Delta(U) \geq 0$, and 
(ii) $\Delta(U) / f(U,q) \leq \Delta(V) / f(V,q)$.
(Since $\Phi(p) > 0$ by (\ref{constr:ind-assumption}), we have that
$q_i < 1$ for each $i$. So, $1/f(U,q)$ and $1/f(V,q)$
are well-defined.)
\end{lemma}

Assuming that Lemma~\ref{lemma:Delta} is true, we 
will now show (\ref{eqn:aim2}); the proof of
Lemma~\ref{lemma:Delta} is given below. We have
\begin{eqnarray*}
\Delta([m]) - \sum_{i} u_i \cdot \Delta(U_i) & = & 
(\Delta([m]) / f([m], q)) \cdot f([m], q) - 
\sum_{i} (\Delta(U_i) / f(U_i,q)) \cdot u_i \cdot
f(U_i, q) \\
& \geq & (\Delta([m]) / f([m], q)) \cdot 
\left[ f([m], q) - 
\sum_i u_i \cdot f(U_i, q) \right] \mbox{~~(by Lemma~\ref{lemma:Delta})} \\
& \geq & 0 \mbox{~~(by (\ref{constr:ind-assumption}) and
(\ref{phip})).} 
\end{eqnarray*}
Thus we have (\ref{eqn:aim2}). 

{\smallskip\noindent\textbf{Proof of Lemma~\ref{lemma:Delta}. }}
It suffices to show the following. Assume $U \not= [m]$; 
suppose $u \in ([m] - U)$ and that
$U' = U \cup \{u\}$.
Assuming by induction on $|U|$ that $\Delta(U) \geq 0$, we
show that $\Delta(U') \geq 0$, and that
$\Delta(U) / f(U,q) \leq \Delta(U') / f(U',q)$. It is easy
to check that this way, we will prove both claims of the lemma. 

The base case of the induction is that $|U| \in \{0,1\}$,
where $\Delta(U) \geq 0$ is directly seen by 
using (\ref{eqn:q-conv-comb}). Suppose 
inductively that $\Delta(U) \geq 0$.
Using the definition of $\Delta(U)$ and the fact
that $f(U', q) = (1 - q_u) f(U, q)$, we have
\begin{eqnarray*}
f(U', q) & = & (1 - q_u) \cdot 
[(1 - p_n) f(U, q') + p_n f(U, q'') - \Delta(U)] \\
& \leq & (1 - (1 - p_n) q_u' - p_n q_u'') \cdot
[(1 - p_n) f(U, q') + p_n f(U, q'')] - (1 - q_u) \cdot \Delta(U),
\end{eqnarray*}
where this last inequality is a consequence of 
(\ref{eqn:q-conv-comb}). Therefore, using the 
definition of $\Delta(U')$ and the facts
$f(U', q') = (1 - q_u') f(U, q')$ and 
$f(U', q'') = (1 - q_u'') f(U, q'')$,
\begin{eqnarray*}
\Delta(U') & = & (1 - p_n) (1 - q_u') f(U, q') +
p_n (1 - q_u'') f(U, q'') - f(U', q) \\
& \geq & (1 - p_n) (1 - q_u') f(U, q') + p_n (1 - q_u'') f(U, q'')
+ \\
& & (1 - q_u) \cdot \Delta(U) - (1 - (1 - p_n) q_u' - p_n q_u'') \cdot
[(1 - p_n) f(U, q') + p_n f(U, q'')] \\
& = & (1 - q_u) \cdot \Delta(U) +
p_n (1-p_n) \cdot (f(U, q'') - f(U, q')) \cdot (q_u' - q_u'') \\
& \geq & (1 - q_u) \cdot \Delta(U) 
\mbox{~~(by (\ref{eqn:qi1-small}))}. 
\end{eqnarray*} 

So, since we assumed that $\Delta(U) \geq 0$, we get
$\Delta(U') \geq 0$; furthermore, we get that
$\Delta(U') \geq (1 - q_u) \cdot \Delta(U)$, which implies
that $\Delta(U') / f(U',q) \geq \Delta(U) / f(U,q)$.
{\hfill $\Box$\medskip} 

\subsection{Proof of Theorem~\ref{theorem:mult-cip}}
\label{sec:deferred-proofs}

(i) Let $E_r \equiv ((Az)_r < b_r)$ be defined w.r.t.\ 
general randomized rounding with parameter $p$; as observed
in Definition~\ref{defn:ch}, $\Pr(E_r) \leq \CH'_r(p)$.
Now if $\CH'_r(p) = 1$ for some $r$, then part (i) is
trivially true; so we assume that 
$\Pr(E_r) \leq \CH'_r(p) < 1$ for all $r \in [m]$.
Defining 
$Z \equiv (Az \geq b) \equiv \bigwedge_{r \in [m]} \overline{E_r}$, 
we get by the FKG inequality that
\[ \Pr(Z) \geq \prod_{r \in [m]} (1 - \Pr(E_r)). \]
Define, for $i = 1, 2, \ldots, \ell$, the ``bad'' event
$\mathcal{E}_i \equiv (c_i^T \cdot z > \lambda_i)$.
Fix any $i$. Our plan is to show that
\begin{equation}
\label{eqn:sss-ith-event}
\Pr(\mathcal{E}_i \bigm| Z) \leq 
\frac{1}{{{\lambda_i} \choose k_i}} \cdot
\sum_{j_1 < j_2 < \cdots < j_{k_i}} 
\left(\prod_{t=1}^{k_i} c_{i,j_t} \cdot p_{j_t} \right) \cdot
\left(\prod_{r \in \mathcal{R}(j_1, j_2, \ldots, j_{k_i})}
(1 - \Pr(E_r))^{-1} \right).
\end{equation}
If we prove (\ref{eqn:sss-ith-event}), then we will be done as
follows. We have
\begin{equation}
\label{eqn:ageqstuff}
\Pr(\mathcal{A}) 
\geq \Pr(Z) \cdot \left(1 - \sum_i \Pr(\mathcal{E}_i \bigm| Z) \right) 
\geq  (\prod_{r \in [m]} (1 - \Pr(E_r))) 
\cdot \left(1 - \sum_i \Pr(\mathcal{E}_i \bigm| Z) \right). 
\end{equation}
Now, the term ``$(\prod_{r \in [m]} (1 - \Pr(E_r)))$'' is a decreasing
function of each of the values $\Pr(E_r)$; so is the lower bound
on ``$-\Pr(\mathcal{E}_i \bigm| Z)$'' obtained from 
(\ref{eqn:sss-ith-event}). Hence, bounds
(\ref{eqn:sss-ith-event}) and (\ref{eqn:ageqstuff}), along with
the bound $\Pr(E_r) \leq \CH'_r(p)$, will complete the proof
of part (i). 

We now prove (\ref{eqn:sss-ith-event}) using 
Theorem~\ref{theorem:sss}(a) and Lemma~\ref{lemma:anti-fkg}.
Recall the symmetric polynomials $S_k$ from (\ref{eqn:S_j}).
Define $Y = 
S_{k_i}(c_{i,1} X_1, c_{i,2} X_2, \ldots , c_{i,n}X_n)/
{{\lambda_i} \choose {k_i}}$.
By Theorem~\ref{theorem:sss}(a), 
$\Pr(\mathcal{E}_i \bigm| Z) \leq \E[Y \bigm| Z]$. Next, the
typical term in $\E[Y \bigm| Z]$ can be upper bounded using
Lemma~\ref{lemma:anti-fkg}:
\begin{eqnarray*}
\E\left[(\prod_{t=1}^{k_i} c_{i,j_t} \cdot X_{j_t}) \bigm| 
\bigwedge_{i=1}^m \overline{E_i}\right]
& \leq & 
\frac{\prod_{t=1}^{k_i} c_{i,j_t} \cdot p_{j_t}}
{\prod_{r \in \mathcal{R}(j_1, j_2, \ldots, j_{k_i})} (1 - \Pr(E_r))}. 
\end{eqnarray*}
Thus we have (\ref{eqn:sss-ith-event}), and the proof of part (i) is
complete. 

\smallskip \noindent
(ii) We have
\begin{equation}
\label{eqn:std-phi0} 
\Phi(p) =
\left[\prod_{r \in [m]} (1 - \CH'_r(p))\right] \cdot
\left(1 - \sum_{i=1}^{\ell} \frac{1}{{{\lambda_i} \choose k_i}} \cdot
\sum_{j_1 < \cdots < j_{k_i}} 
[\prod_{t=1}^{k_i} c_{i,j_t} \cdot p_{j_t}] \cdot
\left(\prod_{r \in \mathcal{R}(j_1, \ldots, j_{k_i})}
\frac{1}{1 - \CH'_r(p)}\right) \right).
\end{equation}
Lemma~\ref{lemma:pmax} shows that under standard randomized rounding, 
$\CH'_r(p) \leq g(B,\alpha) < 1$ for all $r$. 
So, the r.h.s.\ $\kappa$ of (\ref{eqn:std-phi0}) gets lower-bounded
as follows:
\begin{eqnarray*}
\kappa 
& \geq &
(1 - g(B,\alpha))^m \cdot
\left(1 - 
\sum_{i=1}^{\ell} \frac{1}{{{\nu_i(1 + \gamma_i)} \choose k_i}} \cdot
\sum_{j_1 < \cdots < j_{k_i}} 
\left(\prod_{t=1}^{k_i} c_{i,j_t} \cdot p_{j_t} \right) \cdot
[\prod_{r \in \mathcal{R}(j_1, \ldots, j_{k_i})}
(1 - g(B,\alpha))]^{-1}\right) \\
& \geq & \left(1 - g(B,\alpha)\right)^m \cdot
\left(1 - 
\sum_{i=1}^{\ell} \frac{1}{{{\nu_i(1 + \gamma_i)} \choose k_i}} \cdot
\sum_{j_1 < \cdots < j_{k_i}} 
\left(\prod_{t=1}^{k_i} c_{i,j_t} \cdot p_{j_t} \right) \cdot
\left(1 - g(B,\alpha)\right)^{-a k_i}\right) \\
& \geq & \left(1 - g(B,\alpha)\right)^m \cdot
\left(1 - 
\sum_{i=1}^{\ell} 
\frac{{n \choose {k_i}} \cdot (\nu_i/n)^{k_i}}
{{{\nu_i(1 + \gamma_i)} \choose k_i}} \cdot
\left(1 - g(B,\alpha)\right)^{-a k_i}\right),
\end{eqnarray*}
where the last line follows from Theorem~\ref{theorem:sss}(c).
{\hfill $\Box$\medskip} 

\section{Conclusion}
\label{section:concl}
We have presented an extension of the LLL that basically helps
reduce the ``dependency'' much in some settings; we have seen
applications to two families of integer programming problems.
It would be interesting to see how far these ideas can be pushed further.
Two other open problems suggested by this work are: (i) developing a
constructive version of our result for MIPs, and (ii) 
developing a $\poly(n,m)$-time constructive version of
Theorem~\ref{theorem:mult-cip}, as opposed to the 
$\poly(n^{k'},m)$-time constructive version that we present in
\S~\ref{section:construct}. Finally, a very interesting question is
to develop a theory of applications of the LLL that can be made
constructive with (essentially) no loss.

\medskip
\noindent \textbf{Acknowledgements.} This work started while
visiting the Sandia National Laboratories in the summer of 1994; 
I thank Leslie Goldberg and
Z Sweedyk who were involved in the early stages of this work.
I would like to thank \'{E}va Tardos for suggesting the key idea that
helped bootstrap Theorem~\ref{theorem:mip} to get 
Theorem~\ref{theorem:mip-bootstrap}. 
I also thank Noga Alon, Alan Frieze, Tom Leighton, 
Chi-Jen Lu, Alessandro Panconesi, 
Prabhakar Raghavan, Satish Rao, and the SODA 1996
referees for their helpful comments and suggestions.

\end{document}